\documentclass[aps,prb,twocolumn,showpacs]{revtex4}
\usepackage{epsfig}

\newcommand{\bea}{\begin{eqnarray}}
\newcommand{\eea}{\end{eqnarray}}
\newcommand{\be}{\begin{equation}}
\newcommand{\ee}{\end{equation}}

\begin{document}

\title{Disorder, spin-orbit, and interaction effects in dilute ${\rm Ga}_{1-x}{\rm Mn}_x{\rm As}$}
%\author{The authors}
\author{Gregory A. Fiete$^{1,2}$, Gergely Zar\'and$^{1,3}$, Kedar Damle$^{1,4}$, and C. Pascu Moca$^{3}$}

\affiliation{$^1$Department of Physics, Harvard University, Cambridge,
Massachusetts 02138, USA\\
%$^2$Materials Science Division, Argonne National Laboratory, 9700 South Cass Avenue, Argonne, Illinois 60429, USA\\
$^2$Kavli Institute for Theoretical Physics, University of California, Santa Barbara, CA 93106, USA\\
$^3$Research Institute of Physics, Technical University Budapest, Budapest, H-1521, Hungary\\
$^4$Department of
Physics and Astronomy, Rice University, Houston, TX 77005, USA and Department of Theoretical Physics, Tata Institute of Fundamental Research, Homi Bhabha Road, Mumbai 400005 India}

\date{\today}

\begin{abstract}

We derive an effective Hamiltonian for ${\rm Ga}_{1-x}{\rm Mn}_x {\rm
As}$ in the dilute limit, where ${\rm Ga}_{1-x}{\rm Mn}_x {\rm As}$
can be described in terms of spin $F=3/2$ polarons hopping between the
{\rm Mn} sites and coupled to the local {\rm Mn} spins.  We determine
the parameters of our model from microscopic calculations using both a
variational method and an exact diagonalization within the so-called
spherical approximation.  Our approach treats the extremely large
Coulomb interaction in a non-perturbative way, and captures the
effects of strong spin-orbit coupling and Mn positional disorder. We
study the effective Hamiltonian in a mean field and variational
calculation, including the effects of interactions between the holes
at both zero and finite temperature.  We study the resulting magnetic
properties, such as the magnetization and spin disorder manifest in
the generically non-collinear magnetic state. We find a well formed
impurity band fairly well separated from the valence band up to
$x_{\rm active} \lesssim 0.015$ for which finite size scaling studies
of the participation ratios indicate a localization transition, even
in the presence of strong on-site interactions, where $x_{\rm
active}<x_{\rm nom}$ is the fraction of magnetically active Mn.  We study the
localization transition as a function of hole concentration, Mn
positional disorder, and interaction strength between the holes.

\end{abstract}
\maketitle

%75.30.-m Intrinsic properties of magnetically ordered materials
%75.30.Hx Magnetic impurity interactions
%75.47.-m Magnetotransport phenomena
%75.50.Pp Magnetic semiconductors

\section{Introduction}

Recently there has been a surge of interest in the more than 30 year
old field of diluted magnetic semiconductors\cite{reviews} that has
been largely motivated by the potential application of these materials in
spin-based computation\cite{divincenzo,nielson,Awschalom,Zutic}
devices.  In particular, the discovery of {\em ferromagnetism} in
low-temperature molecular beam epitaxy (MBE) grown ${\rm Ga}_{1-x}{\rm
Mn}_x{\rm As}$ has generated renewed interest.\cite{ohno} In this
material Curie temperatures as high as $T_c \approx 160 {\rm K}$ have
been observed.\cite{edmonds}

In this paper we focus on one of the most studied magnetic
semiconductors, ${\rm Ga}_{1-x}{\rm Mn}_x {\rm As}$, though most of
our calculations carry over to other p-doped III-V magnetic
semiconductors.  In ${\rm Ga}_{1-x}{\rm Mn}_x {\rm As}$ substitutional
${\rm Mn}^{2+}$ play a fundamental role: They provide local spin
$S=5/2$ moments and they dope holes into the
lattice.\cite{linnarsson} Since the ${\rm Mn}^{2+}$ ions are
negatively charged compared to ${\rm Ga}^{3+}$, in the very dilute
limit they bind these holes forming an acceptor level with a binding
energy $E_b \approx 112 {\rm meV}$. \cite{linnarsson} As the ${\rm
Mn}$ concentration increases, these acceptor states start to overlap
and form an impurity band, which for even larger ${\rm Mn}$
concentrations merges with the valence band.  Though the actual
concentration at which the impurity band disappears is not known,
according to optical conductivity
measurements,\cite{singley,Singley:prb03} and
ellipsometry\cite{Burch:prb04} this impurity band seems to persist 
up to nominal ${\rm Mn}$ concentrations as high as $x_{\rm nom} \approx
0.05$.  Angle resolved photoemission (ARPES)
data,\cite{asklund,okabayashi,Okabayashi:pe04} scanning tunneling
microscope (STM) results\cite{Grandidier:apl00,Tsuruoka:apl02} and the
fact that even ``metallic'' samples feature a resistivity upturn at
low temperature\cite{vanesch} suggest that for smaller concentrations
(and maybe even for relatively large nominal concentrations) one may
be able to describe ${\rm Ga}_{1-x}{\rm Mn}_x {\rm As}$ in terms of an
impurity band.\cite{fiete,berciu,kennett,Zhou:prb04} While optical
conductivity results\cite{singley,Singley:prb03} and ellipsometry results\cite{Burch:prb04} are suggestive of the presence of an
impurity band in moderately doped samples,\cite{singley,Singley:prb03}
an interpretation based on band to band transitions is also
possible.\cite{Hankiewicz:prb04} We remark, however, that these
materials are {\em extremely dirty}\cite{timm}--the mean free path is
estimated to be of the order of the Fermi wavelength--and therefore it
is not clear if the latter approach is appropriate. Also, ARPES data
indicate that the chemical potential of ``insulating'' samples lies
inside the gap,\cite{asklund} contradicting a band theory-based
interpretation of the optical conductivity data.

A detailed understanding of an impurity band model begins with the
knowledge of a single Mn acceptor state.\cite{Mahadevan:prb04} The
physics of the isolated ${\rm Mn}^{2+}$ + hole system is well
understood: \cite{linnarsson} In the absence of the $\rm{Mn}^{2+}$
core spin, the ground state of the bound hole at the acceptor level is
four-fold degenerate and well described in terms of a $F=3/2$ state.
For most purposes, only the fourfold degenerate $F=3/2$ acceptor levels
need be considered in the dilute limit even in the presence of the
$\rm{Mn}^{2+}$ core spin.  As evidenced by infrared spectroscopy,
\cite{linnarsson} the effect of the $S=5/2$ ${\rm Mn}$ core spin on
the holes is well-described by a simple exchange
Hamiltonian:\cite{reviews}
\begin{equation} 
H_{\rm exch} = G {\vec S} \cdot {\vec F}\;,
\label{eq:Ss}
\end{equation} 
with $G\approx5$ meV. \cite{linnarsson} 

The bound hole (acceptor) states within the $F=3/2$ multiplet are not
Hydrodgenic due to a significant d-wave component of the bound state
wavefunction.\cite{fiete,Tang:prl04,Averkiev:pss94,baldereschi} This
d-wave character ultimately comes from the spin-orbit coupling in GaAs and
has recently been confirmed in the beautiful STM
experiments of Yakunin {\it et. al.}\cite{Yakunin:prl04} The anistropy
of the orbital structure of the wavefunction leads to directionally
dependent hopping of holes between Mn ions, a splitting of the $F=3/2$
level degeneracy, and is expected to strongly influence the magnetic
and transport properties of dilute GaMnAs.\cite{fiete} Here we study
these effects in detail.

One of the main results of this paper is thus the effective
Hamiltonian describing strongly interacting holes hopping from Mn to
Mn. The holes are coupled to the Mn spins via the  exchange interaction 
(\ref{eq:Ss}), where 
\begin{equation}
H^{\rm eff}=H_0^{\rm eff}+H_{\rm int}\;.
\end{equation}
The first part of this Hamiltonian, $H_0^{\rm eff}$, describes the hopping of the holes from Mn to Mn, and the interactions of the Mn acceptor site with the Mn core spin,
\begin{eqnarray} 
H_0^{\rm
eff} & = & \sum_{(i,j)} c^\dagger_{i,\mu} t^{\mu \nu}_{ij} c_{j,\nu} +
\sum_i c^\dagger_{i,\mu} \;( K_i^{\mu\nu} + E_i \;\delta^{\mu\nu})\;
 c_{i,\nu}
 \nonumber \\
& + & G \sum_{i,\mu,\nu} {\vec S}_i \cdot (c_{i,\mu}^\dagger 
{\vec F}^{\mu\nu} \;c_{i,\nu})\;.
\label{eq:hamilt}
\end{eqnarray}
To determine the parameters of (\ref{eq:hamilt}) we shall use the spherical
approximation.\cite{baldereschi} This approach neglects the cubic
symmetry of the lattice, but approximates the band structure rather well
around the top of the valence band at the Gamma point which is most relevant
at the low hole concentrations of interest in the present paper. 
The term  $H_{\rm int}$ accounts for  the on-site interactions of holes 
with each other, and in the spherical approximation,
\begin{equation} 
H_{\rm int} = {U_N\over 2} \sum_i :{{\hat N}_i}^2:
+  {U_F\over 2} \sum_i :{\hat {\vec F}^2}_i:\;.
\label{eq:int}
\end{equation}
The operator $c^\dagger_{i,\nu}$ in the above expressions creates a hole at 
the acceptor level $|F=3/2, F_z = \nu\rangle$ at site $i$, 
${\hat N}_i= \sum_\nu c^\dagger_{i,\nu}c_{i,\nu}$, ${\hat {\vec
F}}_i = \sum_{\mu,\nu} c^\dagger_{i,\mu}{\vec F}^{\mu\nu} c_{i,\nu}$,
and $:...:$ denotes normal ordering. Here ${\vec F}^{\mu\nu}$ is the $\mu \nu$ element of the spin 3/2 matrix.  
The Hubbard interaction strength $U_N$ and the Hunds rule coupling 
$U_F$ in Eq.~(\ref{eq:int}) can be obtained by evaluating
exchange integrals and we find $U_N \approx 2600 $ K and $U_F\approx-51$ K.  

The
presence of nearby $\rm Mn$ sites has three important effects on the
$F=3/2$ acceptor state at any particular $\rm Mn$ site: (i) The Coulomb
potential of the neighboring ${\rm Mn}^{2+}$ ions induces a random
(from the random relative positions of the Mn) {\em shift} $E$ of the
fourfold degenerate states. (ii) Because of the large spin-orbit
coupling in GaAs, the neighboring atoms also generate an
anisotropy $K$ and {\em split} the fourfold {\em degeneracy} of the
$F=3/2$ state into two Kramers degenerate doublets.  (iii) Finally,
the presence of the neighboring ions allow these $F=3/2$ spin
objects to {\em hop} between the ${\rm Mn}$ sites. However, this
hopping $t$ does {\em not conserve the spin} $F$ because of the
spin-orbit coupling. 

To determine the parameters of (\ref{eq:hamilt}),
we performed variational calculations for a dimer of Mn ions
taken to lie along the z-axis where $F_z$ is good quantum
number.  Once the parameters of the dimer is in hand and
the positions of all the Mn are known, the
parameters of the Hamiltonian (\ref{eq:hamilt}) are obtained
by simple spin-3/2 rotations.
%%%%%%%%%%%%%%%%%%%%%%%%%%%%%%%%%%%%%%%%%%%
% GREG: I commented this out after discussing with Gergely.
%%%%%%%%%%%%%%%
% The effective parameters of Eq.~(\ref{eq:hamilt}) are shown in
% Fig.~\ref{fig:couplings} for the case of 2 impurities oriented along
% the z-axis.  For two spins along the z-axis $t_{ij}^{\mu \nu}$ takes
% on two values: $t_{3/2}$ and $t_{1/2}$. These two values differ from
% each other physically because the orbital structure of the
% wavefunction is correlated with its spin when spin-orbit effects are
% included.  For general orientations, rotations like those described
% in Appendix~\ref{app:Nion} need to be made to obtain the $t_{ij}^{\mu
% \nu}$, and the parameters $K_i^{\mu\nu}$ and $E_i$ must be evaluated
% by performing sums over neighboring sites (see
% Appendix~\ref{app:Nion} for details). However, it should be 
% emphasized that all parameters in the effective Hamiltonian $H_0^{\rm
% eff}$ can be determined easily once the positions of the Mn ions and
% the single Mn acceptor state are known.
%%%%%%%%%%%%%%%%%%%%%%%%%%%%%%%%%%%%%%%
%It is important to emphasize that the spherical approximation neglects
%contributions to the energy reflecting the cubic symmetry
%(which typically act as a perturbation on the terms of purely
%spherical symmetry we have retained here) of the
%crystal.\cite{baldereschi} As a result, the important direction in the
%problem is the axis joining any two impurities.  

\begin{figure}[tb]
\begin{center}
\epsfig{figure=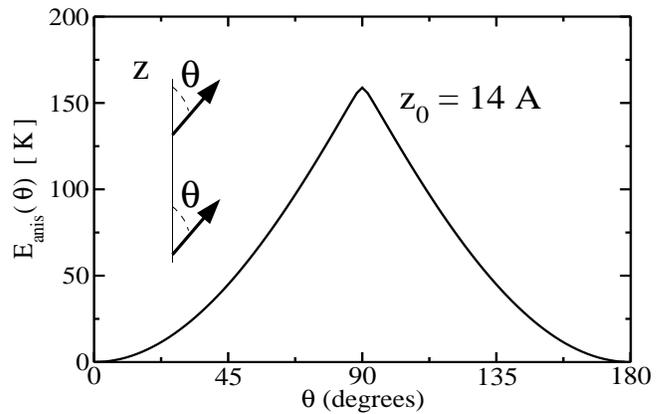, width=8.5 cm,clip}
\end{center}
\vskip0.1cm \caption{\label{fig:easy_axis} Spin orientation
dependence of the ground state energy of one hole on a Mn dimer parallel to the z-axis for a spatial separation of $z_0=14$ \AA. The Hamiltonian (\ref{eq:hamilt}) is diagonalized exactly with the parameters shown in Fig.~\ref{fig:couplings}. As the spins are rotated by an angle $\theta$ away from the z-axis the energy increases and reaches a maximum at $\theta=90 ^o$ before again decreasing.  This indicates the magnetic anisotropy is easy axis.  
}  
\end{figure}

To illustrate the power of the approach, and to better understand the
physical results we obtain from it, consider the simplest case of 2 Mn
impurities and 1 hole.  Diagonalizing Eq.~(\ref{eq:hamilt}) for
different orientations of ${\vec S}_1$ and ${\vec S}_2$ with the
parameters given in Fig.~\ref{fig:couplings} shows that the
magnetization has an easy axis anisotropy (See Fig.~\ref{fig:easy_axis}).  This easy axis anisotropy
immediately leads to frustration among non-collinear Mn positions.

We study the Hamiltonian $H^{\rm eff}=H_0^{\rm eff}+H_{\rm int}$ in
detail using mean field theory when $U_N=U_F=0$ and also with a variational
approach when $U_N,U_F \neq0$. We study the interplay of disorder and
directionally dependent hopping parameters induced by spin-orbit coupling.  
We calculate the temperature dependence of the
magnetization, magnetic anisotropies, the spin distribution functions
measuring the degree of non-collinearity among the spins, the
(impurity band) density of states, and the dependence of the
localization transition on the various parameters of our model.  Our
main results are the following: Qualitatively similar to our earlier
results in the metallic regime,\cite{zarand,Fiete:prb04} we find that
the interplay of disorder and spin-orbit coupling results in (i)
magnetization curves that exhibit linear behavior over a significant
temperature range and (ii) a broad spin distribution function,
implying highly non-collinear magnetic states that result from
spin-orbit induced magnetic anisotropies (iii) within our mean field
and variational calculation we find a well developed impurity band
separated from the valence band for active Mn concentration up to
$x_{\rm active}\lesssim 0.015$ with a localization transition fairly
robust to interactions.

In this paper all Mn concentrations $x$ {\em are the active Mn
concentrations}, i.e., $x=x_{\rm active}$ where active Mn are defined to
be those Mn that contribute to the ferromagnetism of the material.
Interstitial defects with a Mn sitting next to a substitutional Mn may
result in a local singlet formation,\cite{bergqvist} thereby rendering
the two Mn magnetically inactive since they do not contribute to the
ferromagnetism of the material. Thus, the {\em active} Mn concentration is
typically less than the {\em nominal} Mn concentration.  

The interstitial Mn also compensate holes\cite{edmonds,yu} reducing
the number of itinerant holes. In this paper we use the hole fraction
$f$, to relate the hole to the Mn concentration as $N_h=f\,N_{Mn}$
where $N_h$ is the number of holes and $N_{Mn}$ is the number of
active Mn.  Although the precise value of $f$ is not known, typically,
$f=0.1-0.3$.  We thus include the effects of various
compensating
defects,\cite{timm,Timm:rev,Fiete:prb04,potashnik,walukiewicz,Hayashi:apl01}
such as interstitial Mn and As antisites indirectly through the
parameter $f$.

The outline of this paper is the following.  In
Sec.~\ref{sec:variational_calculation} we describe the variational
calculation used to obtain an estimate of the bound state acceptor
wavefunction around a single Mn ion.  In Sec.~\ref{sec:2ion} we use
the variationally obtained wavefunctions to derive and compute the
effective parameters of the Hamiltonian,
Eqs.(\ref{eq:hamilt}) and (\ref{eq:int}), which we then study in
detail in Sec.~\ref{sec:meanfield} using mean field and variational
approaches. Finally, in Sec.~\ref{sec:conclusions} we discuss the main
conclusions of our work. Technical details of our calculations and
various lengthy analytical expressions are relegated to the
appendices.

\section{Variational Calculation of the Baldereschi-Lipari Wavefunctions}
\label{sec:variational_calculation}

In order to study GaMnAs in the dilute limit, we proceed stepwise by
first obtaining bound state (acceptor) wavefunctions in the
single substitutional Mn impurity limit and then using these
wavefunctions to obtain effective parameters of two-ion and $N$-ion
Hamiltonians, details of which are given in Sec.~\ref{sec:2ion}.

\begin{figure}[tb]
\begin{center}
\epsfig{figure=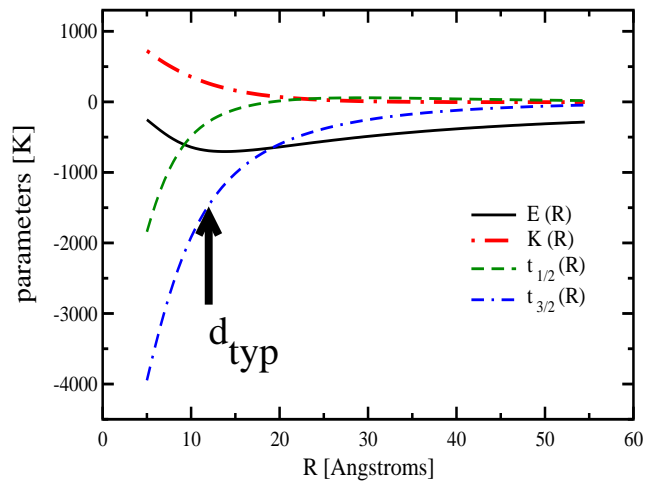, width=8.5 cm,clip}
\end{center}
\vskip0.1cm \caption{\label{fig:couplings} (Color online.) Parameters of the
two-impurity Hamiltonian Eq.~(\ref{eq:two_impurity}) obtained from the
variational study of two Mn ions.  The arrow indicates the typical
Mn-Mn distance, $d_{\rm typ}$, for $x=0.01$ Mn concentration.  }
\end{figure}

We start from the spherical Hamiltonian\cite{baldereschi,zarand,Fiete:prb04} 
\begin{equation}
H_0^{\rm ion} = {\gamma \over 2  m }  \Bigl (p^2 - \mu \sum_{\alpha,\beta}
J_{\alpha\beta} p_{\alpha\beta} \Bigr ) - {e^2\over \epsilon\; r}
+ V_{cc}(r) \;,
\label{eq:app_h01}
\end{equation}
where the central cell correction\cite{yang,bhatta}
\begin{equation} 
\label{eq:cc_corr}
V_{cc}=-V_0e^{-({r\over r_0})^2}\;,
\end{equation}
is used to reproduce the experimentally obtained binding energies, 
and therefore reasonable acceptor
wavefunctions. This affects the parameters given in Fig.~\ref{fig:couplings}
of the effective Hamiltonian (\ref{eq:hamilt}).
Here $r_0$ is a short distance cutoff for the central cell correction and $V_0$
its size.  The primary role of the central cell correction (\ref{eq:cc_corr}) is to take into account atomic interactions in the close vicinity of the
Mn ion. In Eq.~(\ref{eq:app_h01}) 
$\gamma \approx 7.65$ is a mass renormalization parameter, $m$
is the free electron mass, $\mu \approx 0.77$ is the strength of the
spherical spin-orbit coupling in the $j=3/2$ band of GaAs,\cite{baldereschi}
and $\epsilon \approx 10$ is the dielectric constant of GaAs. The
spin-orbit term in Eq.~(\ref{eq:app_h01}) couples the momentum
tensor of the holes $p_{\alpha\beta} = p_\alpha p_\beta -
\delta_{\alpha\beta}\; p^2/3$ to their quadrupolar momentum,
$J_{\alpha\beta} = (j_\alpha j_\beta + j_\beta j_\alpha)/2 -
\delta_{\alpha\beta} \; j(j+1)/3$. This effective Hamiltonian 
gives a relatively accurate value of the hole energy  
in the vicinity of the top of the valence band, but is not very reliable 
for holes with higher energy, since then other states not included in the
derivation of (\ref{eq:app_h01}) will be mixed 
into the acceptor state wave functions. The Hamiltonian (\ref{eq:hamilt}) also does not distinguish between different crystalline directions.
We will discuss the implications
of these features and other shortcomings of the spherical approximation in
the concluding section, Sec.~\ref{sec:conclusions}.

To proceed with the calculation, we note that Eq.~(\ref{eq:app_h01}) can be
made dimensionless by measuring distance in units of the effective
Bohr radius, $a_{\rm eff}=\hbar^2\epsilon\gamma/e^2m=40\AA$, and
taking the corresponding effective Rydberg, $R_{\rm
eff}=e^4m/2\hbar^2\epsilon^2\gamma=15.7$meV, as the energy scale.  In
our calculations we have used $r_0=2.8\AA$ and $V_0=3.0$ eV.  These
values are very close to the numbers used for the central cell
corrections in Ref.[\onlinecite{bhatta}] and Ref.[\onlinecite{yang}].

\begin{figure}[tb]
\begin{center}
\epsfig{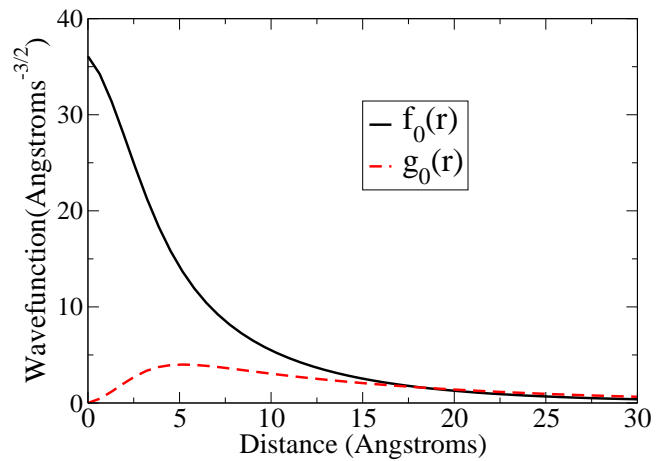}
\end{center}
\vskip0.1cm \caption{\label{fig:fg} (Color online.) Radial wavefunctions obtained from a 
variational calculation for $\mu=0.767$, the relevant value
for GaMnAs. For $r\gtrsim 15\AA$, the typical Mn-Mn distance at $x=0.01$, 
$g_0(r)\approx f_0(r)$. From Eq.~(\ref{eq:Phi}) this means the d-wave
component of the wavefunction is important for typical Mn concentrations 
at $x=0.01$.  It is thus expected that the non-Hydrogenic nature of the 
wavefunction will strongly affect the magnetic and transport properties 
of dilute GaMnAs, a result supported by our numerical calculations presented
in Sec.~\ref{sec:meanfield}.}  
\end{figure}

With the central cell correction, we obtain the correct binding
energy of 112 meV.\cite{linnarsson} However, due to the central cell
correction (\ref{eq:cc_corr}), $a_{\rm eff}$ is no longer a measure of
the spatial extent of the wavefunction as it would be for a purely
Coulomb potential.  Instead, the characteristic length scale is $\sim
10\AA$, as can be seen in Fig.~\ref{fig:fg}.

%\begin{table}%[H] add [H] placement to break table across pages
% \begin{ruledtabular}
% \begin{tabular}{|l|c|c|c|c|c|c|c|r|}
% Source   & $r_0$ &   $a_{\rm eff}$   &   $R_{\rm eff}$   &   $V_0$   
%&  \phantom{n}$E^{us}_{0} $  & $E_{0}^{auth}$\\ \hline
% Ref.~[\onlinecite{yang}] & 2.6 \AA &  40.3 \AA & 17.8 meV & $2.5 $ eV 
%& 74 meV & 124 meV\\
% Ref.~[\onlinecite{bhatta}] & 2.8 \AA & 43.0 \AA & 15.7 meV & 2.7 eV & 102 meV& 
%112 meV\\
%Eq.~(\ref{eq:app_h01}) & 2.8 \AA & 43.0 \AA & 15.7 meV & 3.0 eV & 112 meV & - 
%%%% Lines of table here ending with \\
% \end{tabular}
% \end{ruledtabular}
% \caption{\label{tab:parameters} Numerical values of various authors.
%{\gm Grag: Please, add more text so that the Table + caption is self-contained.}
%} \end{table}

When $\mu \neq 0$ in Eq.~(\ref{eq:app_h01}), the ground state of a hole
bound to an acceptor is no longer a state of zero orbital angular
momentum, $L=0$, since the ``spin-orbit'' term will mix in a $d$-wave,
$L=2$, component.\cite{baldereschi} The ground state wavefunction is
therefore no longer Hydrogenic and hence not spherically
symmetric.\cite{fiete,Averkiev:pss94,Yakunin:prl04} This feature will
lead directly to the appearance of spin-dependent hopping terms 
in Eq.~(\ref{eq:hamilt}).

%different values of $t_{3/2}$ and $t_{1/2}$ in
%Fig.~\ref{fig:couplings} that appear in Eq.~(\ref{eq:hamilt}).  This
%is a consequence of the hole's wavefunction being correlated with its
%spin.

Within the spherical approximation, the total angular momentum 
$\vec F=\vec L+\vec j$
is a constant of the motion and for $\mu \neq 0$ and the ground state 
has $F=3/2$.
The wavefunction for the ground state can then be written
as a sum of an $s$-wave component $f_0$ and a $d$-wave component $g_0$
\begin{eqnarray}
\Phi_{F_z}(\vec{r}) &=& f_0(r)|L=0,j= {3\over 2},F=\frac32,F_z\rangle
\label{eq:Phi}
\nonumber \\ \nonumber \\
 &+&g_0(r)|L=2,j={3\over2},F={3\over2},F_z\rangle\;.
\end{eqnarray}
By acting with the Hamiltonian, Eq.~(\ref{eq:app_h01}), 
on  Eq.~(\ref{eq:Phi}) one
obtains the following set of differential  
equations to be solved for $f_0(r)$ and $g_0(r)$:
\begin{widetext}
\begin{equation}
\pmatrix{-{1\over r}{d^2\over  dr^2}r -{2\over r}+\tilde V_{cc}  
  & %\mu \left({d^2\over  dr^2}+{5\over r} {d\over dr}+{3\over r^2}\right) 
\mu \left({d^2\over  dr^2}-{1\over r} {d\over dr}\right)   
\cr          
\mu \left({d^2\over  dr^2}-{1\over r} {d\over dr}\right)   
&- {1\over r} {d^2\over  dr^2} r +{6 \over r^2}-{2\over r}+\tilde V_{cc}  } 
\pmatrix{f_0(r) \cr g_0(r) \cr} = E_{0} 
\pmatrix{f_0(r) \cr g_0(r) \cr}\;,
\label{eq:system}
\end{equation}
\end{widetext}
where $\tilde V_{cc}\equiv \frac{2 m a_{eff}^2}{\hbar^2 \gamma}V_{cc}$.
In order to solve Eq.~(\ref{eq:system}) we follow the variational approach 
of Ref.~[\onlinecite{baldereschi}] by expanding $f_0$ and $g_0$ as
\bea
f_0(r)& = & \sum_{i=1}^{N} A_i \; f_i(r)\;,
\label{eq:f}
\\
g_0(r)& = & \sum_{i=1}^N B_i\; g_i(r)\;,
\label{eq:g}
\eea
where the $A_i$ and $B_i$ are variational parameters to be determined
and the $f_i(r)$ and $g_i(r)$ are normalized but not orthogonal
basis functions
\bea
%\begin{equation}
%G replaced by alpha_i
%f_i(r)=\frac{2\sqrt{2}(g^{i-1}\alpha)^{3/4}}
%        {\root 4 \of {\pi /2}} e^{- g^{i-1}\alpha r^2}\;,
f_i(r)&=&\frac{2\sqrt{2} \alpha_i^{3/4}}
        {\root 4 \of {\pi /2}} e^{- \alpha_i r^2}\;,
\\
%\end{equation}
%and 
%\begin{equation}
%g_i(r)=r \frac{4\sqrt{2}(g^{i-1}\alpha)^{5/4}}
 %       {\sqrt{3} \root 4 \of {\pi /2}} e^{- g^{i-1}\alpha r^2}\;,
%\end{equation}
g_i(r)&=&r \frac{4\sqrt{2}\alpha_i^{5/4}}
        {\sqrt{3} \root 4 \of {\pi /2}} e^{- \alpha_i r^2}\;,
\eea
with  $\alpha_i=g^{i-1} \alpha$. 
In our computations we have taken $N=21$,
$\alpha=1\times 10^{-2}$, and   $\alpha_{N}=5\times 10^5$ as in 
Ref.~[\onlinecite{baldereschi}], and we also verified that refining the 
basis set resulted in no further improvement. 

To obtain the ground state  wave function, we 
minimize the expectation value of the Hamiltonian  
on the left hand side of Eq.~(\ref{eq:system}). Using the coefficients 
$A_i$ and $B_i$ as variational parameters this involves the 
solution of a simple $2N\times 2N$ eigenvalue problem. One must, however, 
also take into account during this calculation that the states
$f_i$ and $g_i$ are not orthogonal.

%Eq.~(\ref{eq:f})  and Eq.~(\ref{eq:g}) are then 
%substituted into Eq.~(\ref{eq:system}), which leads to a series of guassian
%integrals that make up a (21+21) $\times$ (21+21) matrix.  
The non-orthogonality of the basis set can be  taken into account
through the computation of the overlap matrices, 
$S^{f}_{ij} = \int_0^\infty dr\; r^2 f_i(r)\; f_j(r)$
and
$S^{g}_{ij} = \int_0^\infty dr\; r^2 g_i(r)\; g_j(r)$,
and the transformation of the original problem to a corresponding new 
orthonormal basis following rather standard atomic physics procedures.
The radial functions $f_0(r)$ and $g_0(r)$   obtained in this way
are shown in Fig.~\ref{fig:fg}. Note that while the $s$-wave component $f_0$ 
dominates at short distances, the $d$-wave component $g_0$ becomes appreciable 
for $r> 10 \AA$. This $d$-wave component is ultimately responsible for 
the strong anisotropy of the hopping and effective spin-spin 
interaction.

\begin{figure}[tb]
\begin{center}
\epsfig{figure=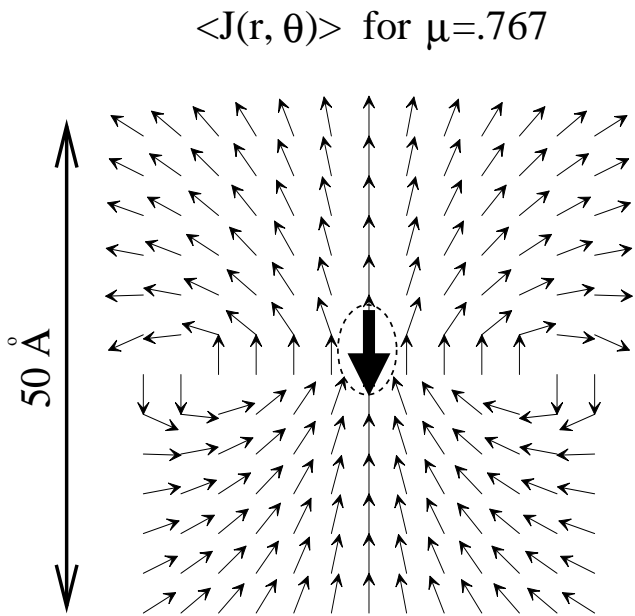, width=8.5cm,clip}
\end{center}
\vskip0.1cm \caption{\label{fig:polar} Polarization of 
a bound hole in the state 
$|F=3/2, F_z = 3/2\rangle$ in $\rm Ga_{1-x}Mn_x As$ around a Mn ion (dark 
arrow pointing downward represents the Mn S=5/2 spin).
Only the direction of the polarization is indicated. The magnitude falls 
off on a scale $\sim 10$~\AA,  as indicated by Fig.~\ref{fig:fg}.}
\end{figure}

Using the radial wavefunctions plotted in Fig.~\ref{fig:fg} one can compute
the expectation value of the  local  spin density, 
$\langle \vec j({\bf r}) \rangle$, around a Mn impurity.
Replacing the Mn spin for a moment with  a classical spin pointing 
downward along the $z$-axis, a bound hole on the acceptor level 
will occupy the state $F_z=3/2$, provided that 
the coupling between the Mn spin and the hole is antiferromagnetic.
The {\em spin direction} (polarization) of this bound hole 
around the impurity is 
shown in Fig.~\ref{fig:polar}.  Note that the polarization direction 
depends on distance and can change sign. Note also that 
in the absence of spin-orbit  coupling, $\mu=0$,  the spin 
polarization of the hole  would be just pointing along the $z$ 
direction, and display RKKY oscillations at larger 
distances (not shown in the figure).
Detailed expressions for the acceptor state spin density are 
given in Appendix~\ref{app:hole_polarization}.

\section{Computing the two-ion  and $N$-ion Hamiltonian}
\label{sec:2ion}

Using the variational wave function obtained in
Sec.~\ref{sec:variational_calculation}, Eq.~(\ref{eq:Phi}), we now
compute the effective parameters of the two-ion hopping Hamiltonian,
Eq.~(\ref{eq:two_impurity}), which will in turn allow us to find
the parameters of the $N$-ion Hamiltonian, Eq.~(\ref{eq:hamilt}), by
using spin-3/2 rotations.

We assume  that we have two impurities separated by a distance $R$.  We take
the quantization axis, $z$, to be along the line joining the two impurities
(ions). Neglecting again the effect of the core Mn spin (for the time 
being), the full Hamiltonian within the spherical approximation 
can be written as
\begin{equation}
H_0^{2-\rm ion} = {\gamma \over 2  m }  \Bigl (p^2 - \mu \sum_{\alpha,\beta}
J_{\alpha\beta} p_{\alpha\beta} \Bigr ) 
+ V_1({\vec r})+ V_2({\vec r}) \;,
\label{eq:Mn_2_full}
\end{equation}
where 
\begin{equation}
V_i({\vec r})= - {e^2\over \epsilon\; \vert \vec{r}-\vec{r}_i\vert}+V_{cc}(\vert \vec{r}-\vec{r}_i\vert),
\label{eq:V1}
\end{equation}
with $\vec r_1$ and $\vec r_2$ the locations of the two
impurities. 

Having computed the single ${\rm Mn}$ hole states, we carried out a
variational calculation to construct the molecular orbitals for a pair
of ${\rm Mn}$ ions in the approximation where we considered only linear 
combinations of the single impurity ground state wave 
functions.\cite{fiete,durst}  
For a pair of Mn spins the full SU(2) symmetry of the 
single impurity model is broken. However, the Hamiltonian 
(\ref{eq:Mn_2_full}) still possesses a cylindrical symmetry, 
corresponding to the conservation of $F_z$. As a consequence, 
the various $F_z$ subspaces decouple, and our task reduces to the 
construction and diagonalization of $2 \times 2$ matrices. 
Furthermore, time reversal symmetry implies that 
the two states with $F_z = \pm 1/2$ and the 
two states with $F_z = \pm 3/2$ remain degenerate.
As a consequence, we find that the two four-fold degenerate 
$F=3/2$ acceptor states of the two Mn impurities are 
split into four Kramers degenerate doublets.
(Details of this calculation are given in 
Appendix~\ref{app:2ion}.) Since for typical Mn distances
these orbitals are well separated from the rest of the spectrum,  
we shall be satisfied by  providing a description of only 
these eight lowest lying states of the ``molecule''. This can 
be achieved by using the following effective Hamiltonian: 
\begin{eqnarray}
&&H^{\rm eff}_{{\rm Mn}-{\rm Mn}} = \sum_\nu t_\nu(R) 
\left( c^\dagger_{1,\nu}c_{2,\nu} + {\rm h.c.} \right ) 
\label{eq:two_impurity} \label{eq:H_2ion} 
\\
&&\phantom{mm}+ \sum_{\scriptstyle i=1,2 \atop \scriptstyle\nu}  \left( K(R) \; \left(\nu^2
-{5\over 4}\right ) + E(R) + E_0\right)  \; c^\dagger_{i,\nu} c_{i,\nu}
\;, \nonumber 
\end{eqnarray}
where $R= |{\vec r_1}-{\vec r_2}|$, $t_\nu$ describes the hopping of 
$F_z=\nu$ holes, $K$ is the splitting of the $F=3/2$ manifold 
of states generated by the presence of the other Mn impurity, 
 and $E$ denotes the  energy shift of the acceptor state 
(at one ion due to the presence of the other ion) with respect to the  
binding energy of an isolated acceptor, $E_0\approx 112 \;{\rm meV}$.  
By time reversal symmetry, the hopping parameters satisfy 
$t_{3/2} = t_{-3/2}$ and $t_{1/2} = t_{-1/2}$.  
All parameters depend only on the
distance $R$ between the two Mn sites (see Fig.~\ref{fig:couplings}).
The most obvious effect of the spin-orbit coupling is that the
hoppings $t_{3/2}$ and $t_{1/2}$ substantially differ from each other;
holes that have their spin aligned with the Mn-Mn bond are more
mobile. As we mentioned in the introduction, this leads to 
an easy axis magnetic anisotropy in the effective spin-spin 
interactions and to non-collinear magnetism.  
As indicated by the arrow in Fig.~\ref{fig:couplings}, 
at the typical Mn-Mn distance for
$x=0.01$, $K$ and $t_{1/2}$ can be entirely neglected compared to 
$E$ and $t_{3/2}$. Therefore, in many cases it is enough to keep only the
latter two terms in the effective Hamiltonian.

So far, we have neglected the interaction between the core Mn spins 
$S$ and the acceptor state. It is known from experiments,\cite{linnarsson} 
that  
the spectrum of an isolated Mn impurity can be very well described by 
a simple exchange Hamiltonian, $H_{\rm exch} = G\; \vec S \cdot \vec F$. 
Furthermore, the separation $\sim 100 \;{\rm meV}$ 
of the acceptor state from other excited  states 
is much  smaller than the experimentally  found exchange coupling
$G\approx 5 \;{\rm meV}$. We can therefore safely treat the exchange field 
of the Mn spin as a {\em perturbation}. We remark at this point that 
the Mn ions are, to a very good approximation, in a $d^5$ state, and 
valence fluctuations on the $d$-levels seem to be rather  small, 
as evidenced by an experimentally observed $g$-factor close to 
2.\cite{linnarsson} In this spirit, we take into account the 
effect of Mn core spins through the following simple term,
\be
H_{\rm exch} =   G \sum_{i=1,2}\sum_{\mu,\nu} {\vec S}_i \cdot 
(c_{i,\mu}^\dagger {\vec F}^{\mu\nu} \;c_{i,\nu})\;.
\ee
Note that in this expression we neglected interactions between the 
core spins and the hole spin on a neighboring Mn acceptor level. This 
approximation 
is certainly justified in the extreme dilute limit, and the above Hamiltonian 
does give a reasonable value for the Curie temperature at the concentrations
we consider.  However,  
 additional terms may be important for a quantitative description 
of GaMnAs.\cite{Mahadevan:prb04}

Finally, let us discuss the hole-hole interaction term, 
Eq.~(\ref{eq:int}). Again, the on site interaction can be greatly 
simplified due to  the presence of SU(2) symmetry within the 
spherical approximation. Since holes are fermions, 
two holes can be placed to the four lowest lying acceptor states
in six different ways. These six states correspond to 
a fivefold degenerate total spin $F = F_1 + F_2 = 2$ two-hole state
and an $F=0$ singlet state. 
The interaction term can be thus written as
\be
H_{\rm int} = U_D \;P_D + U_S \; P_S \;,
\ee
where we introduced the four Fermion operators $P_D$ and 
$P_S$ that project to the $F=2$ and $F=0$ two-hole 
subspaces, respectively. 
With a little algebra we can rewrite these expressions 
in the form Eq.~(\ref{eq:int}), and we can express 
the Hubbard interaction $U_N$
and the Hund's rule coupling $U_F$ in terms of simple 
Coulomb  integrals (see Appendix~\ref{app:onsite_evaluation} for 
details).

%Having computed the effective two-ion Hamiltonian, we can obtain the
%effective $N$-ion Hamiltonian by rotating the $z$-axis along the axis
%joining any 2 impurities.  The details of this transformation
%are given in Appendix~\ref{app:Nion}, and the resulting 
%Hamiltonian, Eq.~(\ref{eq:hamilt}), has already been given in the 
%introduction of this paper.

In the more general case, with three or more impurities, we need to
know how to generalize the Hamiltonian (\ref{eq:H_2ion}) to the
situation where the impurities do not lie along the $z$-axis. We
can derive the parameters of Eq.~(\ref{eq:hamilt}) from the results
of Appendix~\ref{app:2ion} by applying appropriate rotations.

This can be achieved as follows. Assume that we have two Mn impurities
at positions ${\vec r}_1$ and ${\vec r}_2$. It is trivial to write 
the hopping part of the Hamiltonian if we quantize the spin of the holes 
along the unit vector ${\vec n} = (\sin(\theta)\cos(\phi), 
\sin(\theta)\sin(\phi), \cos(\theta)) $ 
connecting 
 ${\vec r}_1$ and ${\vec r}_2$. Denoting the eigenvalues of 
${\vec F} \cdot {\vec n}$ by $\tilde \nu$, we can write the hopping 
part of the Hamiltonian in the simple form
\be
H^{\rm hop}_{{\rm Mn}-{\rm Mn}} = \sum_{\tilde \nu} t_{\tilde \nu}(R) 
\left( c^\dagger_{1,\tilde \nu}c_{2,\tilde \nu} + {\rm h.c.} \right )\;,
\ee 
where $c^\dagger_{i,\tilde \nu}$ creates a hole at site $i$ with 
${\vec F} \cdot {\vec n} = \tilde \nu$, 
and $R$ denotes the separation between the two ions.
We need to re-express
this Hamiltonian in terms of operators that create holes with 
$F$ quantized along the $z$-axis. This can be simply achieved by noticing 
that these two sets of operators are related by a unitary transformation:
\be
c^\dagger_{\tilde \nu} = \sum_{\nu} [U({\vec n})]_{ \nu, \tilde \nu} 
\;c^\dagger_{\nu}\;,
\ee
where $U$ is just the usual spin $3/2$ rotation matrix:
\be
U({\vec n}) = e^{i \phi F_z} e^{i \theta F_y} \;.
\ee
Making use of this transformation we can rewrite the hopping term 
in this standard basis as 
\be
H^{\rm hop}_{{\rm Mn}-{\rm Mn}} = \sum_{\nu, \nu'} 
\left(
t_{12}^{\nu\nu'}  c^\dagger_{1,\nu}c_{2,\nu'} + 
{\rm h.c.} 
\right )\;,
\ee 
where the hopping matrix is simply given by
\be 
t_{12}^{\nu\nu'}
 = \sum_{\tilde \nu}
% \bigl[e^{i \phi F_z} e^{i \theta F_y}  \bigr]_{\nu \tilde \nu}
\bigl[U({\vec n})\bigr]_{\nu \tilde \nu}
\; t_{\tilde \nu}(R) \;
\bigl[U^\dagger({\vec n}) \bigr]_{\tilde \nu \nu'}\;.
\ee
It is much simpler to generalize the spin splitting term 
$\sim K$ which can trivially be written as
$$
H_{\rm Mn-Mn}^{\rm anis} = \sum_{i=1,2}
K(R) c^\dagger_{i,\nu} 
\bigl[(\vec n\cdot\vec F)^2_{\nu\nu'}-\frac54 \delta_{\nu\nu'}\bigr]
c_{i,\nu'}\;.
$$
Finally, the energy shift term is manifestly invariant with respect to the 
spin quantization axis,
\be
H_{\rm Mn-Mn}^{\rm shift} =  E(R)  \sum_{i=1,2}\sum_{\nu}
 \; c^\dagger_{i,\nu} c_{i,\nu}\;.
\ee

For a finite number of ions  the above perturbations add up in a tight binding 
approach, leading to the effective Hamiltonian (\ref{eq:hamilt})
with 
\bea 
K_i^{\mu\nu} &=& \sum_{j\ne i} K(R_{ij}) 
\bigl[({\vec n}_{ij}\cdot\vec F)^2-\frac54\bigr] _{\mu\nu}\;,\\
t_{ij}^{\nu\nu'}
& =& \sum_{\tilde \nu}
\bigl[U({\vec n_{ij}})\bigr]_{\nu \tilde \nu}
\; t_{\tilde \nu}(R_{ij}) \;
\bigl[U^\dagger({\vec n_{ij}}) \bigr]_{\tilde \nu \nu'}\;,
\eea
and 
\be
E_i = E_0 + \sum_{j\ne i} E(R_{ij})\;.
\label{eq:E_i}
\ee
We remark here that for large distances $E(R)$ scales as 
$1/R$ and therefore, strictly speaking, the latter 
sum is not convergent. This unphysical result of our approach,
 which does not take into account 
{\em screening}, 
can be remedied in our calculation by introducing an exponential cutoff 
of the order of the Fermi wavelength in Eq.~(\ref{eq:E_i}).

This completes the derivation of the parameters of the general Hamiltonian
(\ref{eq:hamilt}), aside from the on-site interaction described in 
Appendix~\ref{app:onsite_evaluation}.

\section{Mean Field and Variational Study of the Effective Hamiltonian}
\label{sec:meanfield}

In this section we study the effective Hamiltonian (\ref{eq:hamilt})
in a mean field theory\cite{Kennett:prb02} and within a variational
calculation when the interaction (\ref{eq:int}) is 
also included.\cite{Xu:prl05,Timm:rev}  Throughout this section we shall 
treat the Mn core spins as {\em classical variables}.
Our main goal is to study the interplay of disorder in the Mn positions and
spin-orbit coupling of the GaAs host 
on the magnetic  properties of dilute GaMnAs. 
 Due to spin-orbit effects in the GaAs
host, the effective Mn spin-spin interactions are expected to be
anisotropic,\cite{zarand,Timm:condmat04} and these anisotropies are
expected to be greater for smaller concentrations of Mn ions and
holes.\cite{fiete,Fiete:prb04}

\subsection{Computational methods}

Most of our calculations have been performed in the 
absence of the interaction term, $H_{\rm int}$, 
where we used a simple mean field treatment of the spins.\cite{berciu}
In this approximation one has to solve a set of 
equations self-consistently. 

The first one of these equations just expresses the fact that polarization 
of the  impurity spin $S_i$ is generated by the effective
 field $G \langle \vec F_i\rangle$ generated  in turn 
by the polarization of the
hole spins:
\bea
\langle \vec S_i\rangle  = S\;  {\vec \alpha_i\over \alpha_i}
\bigl( {\rm coth}(\alpha_i) - {1\over\alpha_i}\bigr)\;,
\phantom{nn}
\vec \alpha_i = {G\over T} S \langle \vec F_i\rangle \;.
\eea
Te second equation gives  the 
effective Hamiltonian of the holes that must be used to compute 
the thermodynamical  average $ \langle \vec F_i\rangle$,
\bea 
H_0^{\rm eff}\to \tilde H_0^{\rm eff} &= &
 \sum_{(i,j)} c^\dagger_{i,\mu} t^{\mu \nu}_{ij} c_{j,\nu} 
 \nonumber 
\\
& +&
\sum_i c^\dagger_{i,\mu} \;( K_i^{\mu\nu} + E_i \;\delta^{\mu\nu})\;
 c_{i,\nu} \nonumber 
\\
& + & G \sum_{i,\mu,\nu} \langle {\vec S}_i\rangle  \cdot (c_{i,\mu}^\dagger 
{\vec F}^{\mu\nu} \;c_{i,\nu})\;.
\label{eq:MF_effect}
\eea
Here the last term simply expresses that a non-zero average of 
$ \langle {\vec S}_i\rangle $ acts as a local field on the holes and tries 
to polarize them. Note that the latter Hamiltonian is quadratic. 
Therefore, once it is diagonalized
and its eigenfunctions are constructed, we can construct the corresponding 
density matrix and compute the finite temperature expectation 
values  $ \langle \vec F_i\rangle$ in a relatively straightforward way,
and thus solve the above equations iteratively. 

Although the Hubbard coupling $U\equiv U_N$ in Eq.~(\ref{eq:int}) is rather 
large, at small hole fractions two holes overlap with a small 
probability, and therefore this interaction term is not expected 
to play a crucial role.\cite{berciu} To verify these expectations, 
we carried out  calculations for the 
interacting Hamiltonian with $U_N\ne0$ at $T=0$ temperature. 
The Hund's rule coupling  $U_F$ being rather  small, we 
neglected this interaction term throughout these computations.

A full Hartree-Fock treatment of $U\equiv U_N$ is cumbersome: it requires
the self-consistent determination  $18$ effective fields
at each site, and we typically experienced serious convergence problems 
while trying to determine these fields. 
However, the essential
effects of the interaction term (\ref{eq:int}) can be captured by a
simpler approach that retains the variational character of
Hartree-Fock theory. In such a variational approach, we replace the
interacting Hamiltonian $H(\vec{S}_{i})$ by a non-interacting Hamiltonian
\begin{equation}
H_{var}(\mu_i, \vec{h}_i, \vec{S}_{i}) \equiv H_0^{\rm eff}(\{\vec{S}_{i}\}) -
\sum_i \mu_i \hat{N}_i + \sum_i \vec{h}_i
\cdot \vec{F}_i\;,
\label{eq:hamilt2} 
\end{equation}
where the variational parameters $\vec{h}_i$ and
$\mu_i$ are {\em numerically} determined by minimizing (for fixed
$\{S_i\}$)  the expectation value of the full Hamiltonian 
$\langle \phi_{\rm var}|  H|  \phi_{\rm var} \rangle _{var}$, 
Eqs.~(\ref{eq:hamilt}) and (\ref{eq:int}), using 
the ground state $ \phi_{\rm var}$ 
of $H_{var}$. 

A $T=0$ minimization with respect to the spins  $\vec{S}_i$
leads to the condition that  the spins must  
be aligned  anti-parallel to the expectation values of the
corresponding $\vec{F}_i$ in this variational ground state. 
Therefore, after finding the expectation values 
 $\langle \vec{F}_i\rangle $ in the variational ground state
for a given spin configuration $\{\vec S_i\}$, we 
generate a new spin configuration by aligning all spins 
anti-parallel to the   $\langle \vec{F}_i\rangle $'s.
This
procedure is then iterated with the new values of $\vec{S}_i$ to
obtain a self-consistent variational solution that includes the effect
of interactions.  In practice, even this restricted approximation is
very time-consuming because the minimization of the variational energy
at fixed $\vec{S}_i$ is computationally expensive. The procedure
outlined above could therefore be carried out for only very small
sample sizes. Below, we therefore 
present results obtained through  a restricted variational approach that
only uses the variational parameters $\mu_i$ at each site. For
satisfactory convergence of the variational energy minimization step,
we slowly crank up $U_N$ from $0$ to its final value in steps of
$\Delta U= 10\;{\rm K}$.  

%This approach can easily be generalized to finite temperatures.

\begin{figure}[h]
\begin{center}
\epsfig{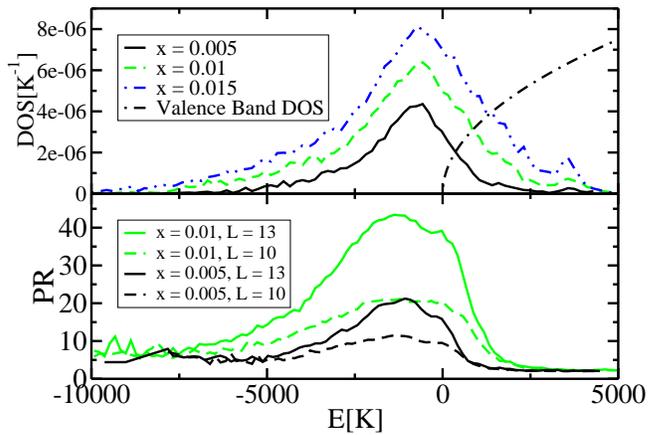}
\end{center}
\vskip0.1cm \caption{\label{fig:dos_x} (Color online.) Top: The dependence of the density of 
states on doping $x$ for $L=10\,a_{\rm lat}$, $f=0.50$, and $t_{MC}=0$. 
Data is the average of 50 sample realizations. The DOS is normalized to 
the volume of a unit cell,
so the total number of states is proportional to $x$.  The half-width of the
impurity band ranges from $1000-2500$ K and is centered around -1100 K, the
binding energy of a hole at an isolated Mn.  The shape of the density
of states changes little with the Mn concentration, $x$, over the range of
values shown.  The value of the Fermi energy
is $\approx -5000$ K.  For
comparison, the valence band density of states is also shown. Bottom: The dependence of the participation ratio on doping $x$.  Data is averaged over 50 samples.  Larger samples have larger values of the PR for delocalized states, while for localized states the PR is independent of
system size for fixed $x$, $t_{MC}=0$.  The energy value that separates 
$L$-dependent PRs from $L$-independent PRs is the mobility edge.  This
depends on $x$ and is larger for larger $x$.  For the disordered samples here,
the mobility edge is not too sharp and lies in the tail of the 
density of states.
}
\end{figure}

\begin{figure}[h]
\begin{center}
\epsfig{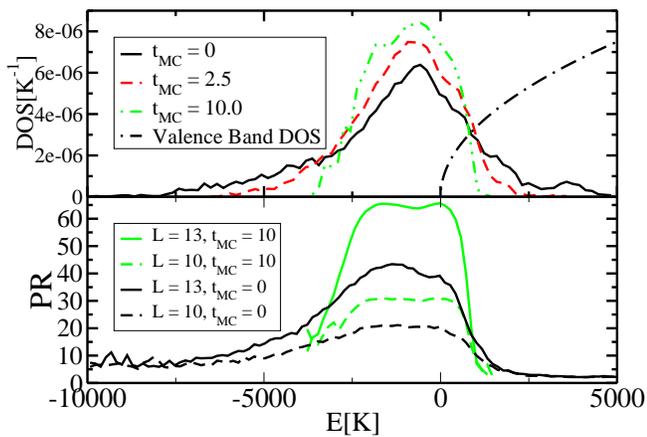}
\end{center}
\vskip0.1cm \caption{\label{fig:dos_MC} (Color online.) Top:
The dependence of the density of states on Monte Carlo time for $U=0$,
$x=0.01$, $f=0.30$, and $L=10\,a_{\rm lat}$.  Data is the average of 50 sample realizations.
MC time $t_{MC}=0$ means that the Mn positions
are completely random; for $t_{MC}=10$ the Mn ions form a nearly BCC
lattice with a few point defects.  The main effect of disorder is thus to
broaden the impurity band. The width of the impurity band
is proportional to the value of the dominant hopping parameter, $t_{3/2}$, at 
typical Mn separations as can be seen from Fig.~\ref{fig:couplings}.
Bottom: The dependence of the 
PR on the MC time for $x=0.01$ and $f=0.30$. Data is averaged over 50 samples. 
The mobility edge also moves up to higher 
energy values for more ordered Mn configurations and nearly all
states become delocalized.
}
\end{figure}

\begin{figure}[h]
\begin{center}
\epsfig{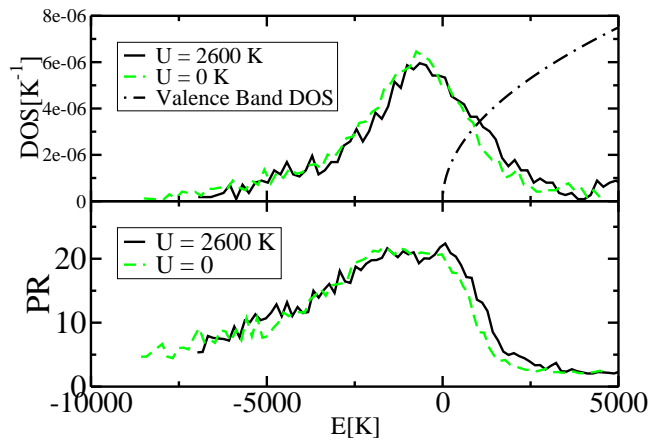}
\end{center}
\vskip0.1cm \caption{\label{fig:kedar_dosdata} (Color online.) Top: The 
dependence of the 
density of states on the on-site interactions, $U=2600$ K, for $x=0.01$, 
$f=0.30$, and $t_{MC}$. The effect
of interactions on the DOS is minimal: The overall shape remains 
largely unchanged by the interactions.  Bottom: The dependence 
of the 
PR on the on-site interactions, $U$, for $x=0.01$, $f=0.30$, $L=10\,a_{\rm lat}$, and $t_{MC}=0$. Data is averaged over 10 samples.  The behavior of
the participation ratio follows roughly that of the density of states
shown in Top; There is little shape change compared to the non-interacting case.}
\end{figure}

In our calculations we considered samples of fixed size $L=10\;a_{\rm
lat}$ and $L=13 \;a_{\rm lat}$ where $a_{\rm lat}$ is the length of
the edge of the FCC unit cell.  The effective Hamiltonian
(\ref{eq:hamilt}) and (\ref{eq:int}) is only expected to be valid in
the very dilute limit of Ga$_{1-x}$Mn$_x$As, so we considered only 
active Mn concentrations
$x=0.005,0.01$ and 0.015. 
The validity of our approach can be checked
post-facto by noting that the high energy tail of the impurity band
has fairly small overlap with the valence band density of states
for these concentrations,  as
seen in Figs.~\ref{fig:dos_x}-\ref{fig:kedar_dosdata}.  
Compensation effects have been taken into account through the 
hole fraction parameter, $f$. Although this parameter is not precisely known 
for low-concentration samples, we used the values 
$f=0.1-0.3$, typically assumed in the literature.

In order to control the amount of disorder,
we introduced a screened
Coulomb repulsion between the Mn ions and let them relax using zero temperature
Monte Carlo (MC) simulations as described in Ref.~[\onlinecite{Fiete:prb04}]. 
We found that the Mn ions relax to their long time
configuration approximately exponentially fast with a characteristic
relaxation time $t_{MC}^{\rm relax}\approx 2.5$, and that for
 long times the Mn ions form a regular BCC
lattice with some point defects. Such calculations are not meant to model
real defect correlations\cite{timm,Timm:rev} in GaMnAs, but rather to help 
understand
how the disorder in the material affects its physical properties, especially
when random ion positions are important as they are for small $x$ and
small carrier concentrations.\cite{Fiete:prb04}

Once the Mn positions are fixed in a given instance, the mean field
equations derived from (\ref{eq:hamilt}) are solved self-consistently.
\cite{Kennett:prb02}  We usually start 
the iterative procedure from a configuration where all 
Mn spins are aligned in one direction.  We used periodic boundary conditions and
implemented a short distance cutoff in the hopping parameters of
Eq.~(\ref{eq:hamilt}) which corresponds to about 8 neighbors for each
Mn. The use of this cutoff is justified by the observation that our
molecular orbital calculations are only appropriate for ``nearest
neighbor'' ion pairs, and in reality, holes can not hop directly over
the first ``shell'' of ions.

\begin{figure}[h]
\begin{center}
\epsfig{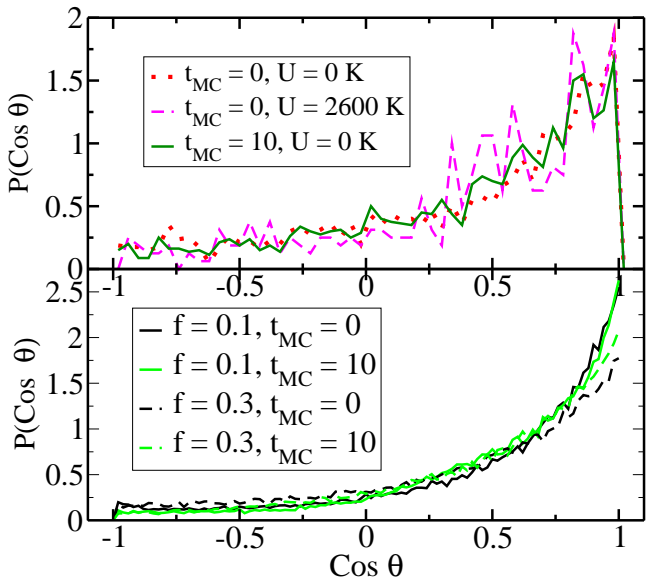}
\end{center}
\vskip0.1cm \caption{\label{fig:spin_distribution_L10} (Color online.)  Top: 
The dependence of the 
spin distribution function, $P(\cos(\theta))$, on the on-site interaction, 
$U=U_N$, and
the Monte Carlo time for $L=10\,a_{\rm lat}$, $x=0.01$, and $f=0.30$. 
We
 averaged over 10 samples. 
Even with interactions and at large Monte Carlo times (small
disorder) the spin distribution function remains broad.  This is consistent 
with the strong reduction of
the saturation magnetization ($\sim 60\%$) observed in our calculations, 
independent of Monte Carlo time. Bottom:
Dependence of the spin distribution on the hole fraction $f$, $U=0$, 
obtained after averaging over 
100 samples.}
\end{figure}

\begin{figure}[h]
\begin{center}
\epsfig{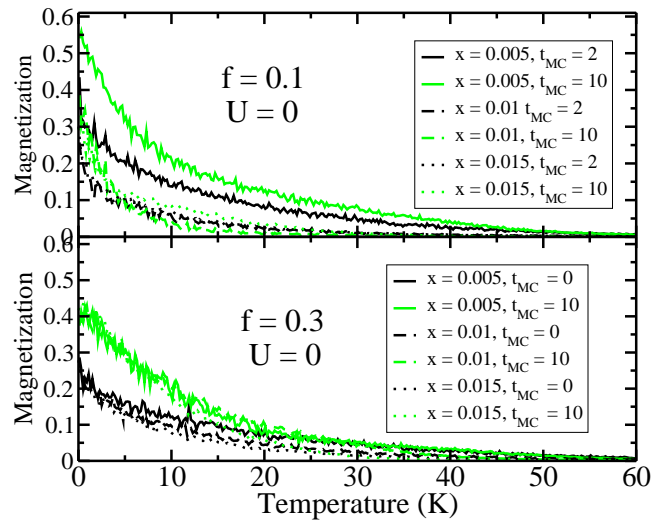}
\end{center}
\vskip0.1cm \caption{\label{fig:magnetization} (Color online.)  Magnetization as a function of temperature, Mn concentration $x$, and Monte Carlo time $t_{MC}$ for different hole fractions $f$. Here $L=10a_{\rm lat}$, $U=0$, and 100 samples are averaged over. Top: Hole fraction $f=0.1$. Bottom: Hole fraction $f=0.3$. In both cases, as the Monte Carlo time increases, for fixed $x$, the saturation magnetization at zero temperature increases.  For both values of $f$ the curves remain linear over a fairly wide temperature range, much the same as for experimentally measured curves for unannealed GaMnAs. The saturation magnetization never reaches more than $\sim 60\%$ of the fully saturated value.  This is consistent with the wide spin distribution shown in Fig.~\ref{fig:spin_distribution_L10} and indicates that the ferromagnetism is non-collinear.
}
\end{figure}

\begin{figure}[h]
\begin{center}
\epsfig{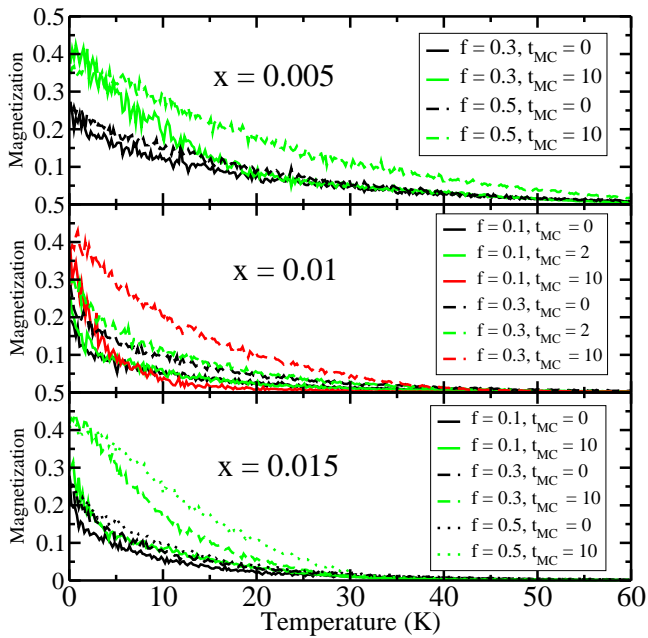}
\end{center}
\vskip0.1cm \caption{\label{fig:magnetization_x} (Color online.) Magnetization as a function of temperature, hole fraction $f$, and Monte Carlo time $t_{MC}$ for different Mn concentrations $x$. Here $L=10\,a_{\rm lat}$, $U=0$ and 100 samples are averaged over.  Compare with Fig.~\ref{fig:magnetization}. Top: Mn concentration $x=0.005$. Middle: Mn concentration $x=0.01$. Bottom: Mn concentration $x=0.015$.  The general trend is the same as in Fig.~\ref{fig:magnetization}: Longer Monte Carlo times lead to larger zero temperature magnetizations.  The saturation magnetization is roughly independent of Mn concentration $x$.
}
\end{figure}

\subsection{Results}

\subsubsection{Magnetization}

Similar to the metallic case within the spherical
approximation,\cite{zarand} we find a ferromagnetic state with a
largely reduced magnetization, $|\langle \vec \Omega_i\rangle|\approx
0.4$ for $L= 10\, a_{\rm lat}$. (See Figs.~\ref{fig:magnetization} 
and \ref{fig:magnetization_x}.)  We find that this reduction is largely
due to spin-orbit coupling, and that 
$\cos(\theta_i)={\vec \Omega_i} \cdot{\vec n}$, 
(where ${\vec n}$ is the direction of the
ground state magnetization vector) has a broad
distribution, $P(\cos(\theta))$, quantitatively similar to 
earlier results obtained in the 
metallic case using the
4-band spherical approximation in the completely disordered
case.\cite{zarand} The
interaction Hamiltonian (\ref{eq:int}) appears to have a
negligible effect on the spin distribution.  Also, relaxing the Mn
impurities to form a regular BCC lattice as described above appears to
have little impact on the spin distribution. We checked that this
result is valid at least for $f=0.1-0.3$.  This is {\em qualitatively}
different from the metallic case which showed a significant sharpening
of the distribution function as the Mn positions became more ordered,
and a corresponding increase of the saturation magnetization to an
almost fully polarized state.\cite{Fiete:prb04}

The magnetization for $U=0$ is shown in Figs.~\ref{fig:magnetization}
and \ref{fig:magnetization_x}.  The curves indicate that the system
never reaches the fully polarized state, even for long Monte Carlo
times.  However, as the disorder is reduced the saturation
magnetization increases from $\approx 20-25\%$ to $40-50\%$. The
magnetization curves exhibit linear behavior over a large temperature
range, qualitatively similar to experiments on disordered samples.

Unfortunately, since the numerical calculations are rather demanding, 
we could not perform a proper finite size scaling analysis. 
Therefore, while our calculations suggest that the ground state of our model is 
ferromagnetic, we cannot exclude the possibility of a paramagnetic or spin glass 
state for these small concentrations.

\subsubsection{Density of States}

We compute the DOS from the Hamiltonian (\ref{eq:hamilt})
and in the interacting case $H_{var}$.  The results are shown in
Figs.~\ref{fig:dos_x}-\ref{fig:kedar_dosdata}.

Fig.~\ref{fig:dos_x} shows the dependence of the DOS on doping $x$ for
fixed MC time and $U_N=U_F=0$.  The total number of states is
proportional to $x$.  The overall shape is fairly independent of $x$,
over the range of $x$ considered here, which shows a peak near the
binding energy, $E_b\approx -1100$K, of the isolated Mn+hole system
and a half-width of 0.1-0.25 eV.  The impurity band slightly overlaps
the valence band DOS. However, comparison with the valence hole
density of states suggests that at concentrations $x\lesssim 0.015$ a
well-formed impurity band may still be present, and it might persist
to higher concentrations.  Indeed, this scenario seems to be supported
by many
experiments.\cite{singley,asklund,Singley:prb03,Burch:prb04,okabayashi,Okabayashi:pe04,Grandidier:apl00,Tsuruoka:apl02}

Fig.~\ref{fig:dos_MC} shows the dependence of the DOS on MC time for fixed
$x$.  For $t_{MC}=0$ the Mn ions are completely random while for $t_{MC}=10$
the Mn ions form a nearly perfect BCC lattice with a few point defects.  The
main effect of disorder, mostly due to the random Coulomb shift of $E_i$ in
Eq.~(\ref{eq:E_i}), is to broaden the impurity band DOS.  In the
ordered case, the width of the impurity band is determined by the value
of the dominant hopping parameter, $t_{3/2}$ at the typical Mn separations.

Fig.~\ref{fig:kedar_dosdata} shows the effects of the interactions 
on the DOS.  Within the variational calculation, the absolute scale of the
quasi-particle energies is not given. However, as shown in 
Fig.~\ref{fig:kedar_dosdata}, the overall shape of the single particle density of states and the energy-dependent participation ratio are almost identical 
to what we found in our calculations performed for the non-interacting 
model. 

In order to gain information 
on transport properties of the holes, we turn to an analysis 
of another quantity,
the participation ratio, from which finite size scaling will be able to
tell us which states of the impurity band are localized and which states
are delocalized in the impurity band.

\subsubsection{Participation Ratios}  

The participation ratio, ${\rm PR}= [ \sum_i (\sum_\alpha
\vert\psi_{i\alpha}\vert^2)^2]^{-1}$, measures the degree to which
wavefunctions are localized.  If states are completely delocalized, the
single particle wavefunction $\psi_{i\alpha}$ will be spread equally over all sites making
the PR system-size dependent because the wavefunction must be normalized
to unity.  Thus, the PR grows with system size for
delocalized states while the it remains ${\cal O}$(1) in the thermodynamic
limit for localized states. 
% The PR can thus be used to study the localization
%transition within the impurity band model resulting from the Hamiltonian
%(\ref{eq:hamilt}) and (\ref{eq:int}) studied here.

Fig.~\ref{fig:dos_x} shows the dependence of the PR on $x$ and system
size $L$ for fixed disorder.  Larger samples have larger values of the
PR for delocalized states while for localized states the PR is $L$
independent.  The energy joining the two regimes is the mobility edge.
It is impossible to determine the precise position of the 
mobility edge from our numerics, but
in both cases, the Fermi energy apparently lies in the region of {\em
delocalized} states, indicating a localization transition in the
impurity band itself.

Fig.~\ref{fig:dos_MC} shows the dependence of the PR on disorder. 
For small disorder, nearly all states
become delocalized and similar to the the disordered case 
the localization transition occurs in the impurity band.

Fig.~\ref{fig:kedar_dosdata} shows the dependence of the PR on the
on-site interactions in Eq.~(\ref{eq:int}).  The behavior of the PR
ratio roughly follows that of the DOS shown in
Fig.~\ref{fig:kedar_dosdata}: There is little shape change with
the interactions, and the
result looks very similar to the
non-interacting case.  Thus, the relation between the mobility edge
and the Fermi energy remains essentially unchanged implying that the
localization transition is robust to reasonable on-site interactions.

To summarize the result of this section, we find that the chemical
potential lies deep ($\sim 0.5-0.7\, {\rm eV}$) inside the gap. From
the PR data, it appears that the chemical potential is in the vicinity
of the mobility edge, a regime where our model is probably more
reliable.  This suggests that the localization phase transition in
${\rm Ga}_{1-x}{\rm Mn}_x {\rm As}$ could happen inside the impurity
band and that the ferromagnetic phase for smaller Mn concentrations is
governed by localized hole
states. \cite{fiete,berciu,Berciu:prb04,timm,yang,Chattopadhyay:prl01,Galitski:prl04}

\section{Conclusions}
\label{sec:conclusions}

Starting with a single Mn acceptor state in GaMnAs, we derived an
effective Hamiltonian for ${\rm Ga}_{1-x}{\rm Mn}_x {\rm As}$ valid in
the dilute limit, where ${\rm Ga}_{1-x}{\rm Mn}_x {\rm As}$ can be
described in terms of spin $F=3/2$ polarons hopping between the {\rm
Mn} sites and coupled to the local {\rm Mn} spins.  We estimated the
parameters of this model from microscopic calculations using both a
variational approach and an exact diagonalization for a pair of Mn ions
within the spherical approximation. Our approach treats
the extremely large Coulomb interaction in a non-perturbative way, and
captures the effects of strong spin-orbit coupling, and disorder. 
We find that due to the large spin-orbit coupling of GaAs, the 
hopping matrix elements of the polarons depend on their 
{\em spin direction}.

We studied the above effective Hamiltonianon using mean field and variational
methods, also including the effects of interactions between the holes.
We find that the spin-dependent hopping generates frustration  and is  
ultimately responsible for the formation of a non-collinear magnetic state  
for small active Mn concentrations. The existence of such non-collinear ground states
is indeed supported by experiments, where a substantial increase in the 
remanent magnetization is found upon the application of a relatively 
small magnetic field in some unannealed samples.\cite{Ku:apl03}

Our calculations also support  the existence of an impurity band, and a 
metal-insulator phase transition inside this impurity band
for these small concentrations of active Mn ions, in agreement with 
angle resolved photoemission (ARPES) data,\cite{asklund,okabayashi,Okabayashi:pe04} 
scanning tunneling microscope (STM) results,\cite{Grandidier:apl00,Tsuruoka:apl02}
and  optical conductivity measurements.\cite{singley,Singley:prb03} 

The main advantage of our approach  is that it provides a clear
description of the most important physical ingredients needed to
describe dilute ${\rm Ga}_{1-x}{\rm Mn}_x {\rm As}$, while it treats the 
extremely large Coulomb potential of charged  substitutional Mn ions non-perturbatively.
While the resulting effective Hamiltonian given by  Eqs.~(\ref{eq:hamilt}) and (\ref{eq:int}) 
is relatively simple, it captures  many of the physical properties of 
 ${\rm Ga}_{1-x}{\rm Mn}_x {\rm As}$, and can serve as a starting point for 
field theoretical computations of other physical quantities of interest 
such as  optical conductivity, spin wave relaxation rate, conductivity 
or (anomalous) Hall resistance.

Though the parameters of our effective Hamiltonian have been determined 
from microscopic model calculations, they are  only approximate:  while the spherical
approximation  used is able to reproduce the spectrum of a single
acceptor rather well, it certainly  overestimates the effect of spin-orbit
coupling and the width of the impurity band. A direct comparison of the parameters in
Fig.~\ref{fig:couplings} with those obtained from a more accurate
six-band model calculation shows some important quantitative differences.\cite{Pawel}
This comparison reveals that while for $r \gtrsim 13 \AA$ the effective
Hamiltonian (\ref{eq:hamilt}) is indeed a good approximation in form,
the hopping parameters are smaller by a factor of two compared to the
ones obtained from the six band model variational calculation.  Moreover, for $r
\lesssim 13 \AA$, the six-band model gives $t_{3/2} \approx t_{1/2}$,
suggesting that spin anisotropy is much smaller than that obtained from
the spherical model. Furthermore, for shorter Mn separations the effective model turns out 
to be a rather poor approximation.\cite{Pawel}

%ore realistic {\it ab initio}
%calculations would be needed to give a quantitative answer concerning
%the role of the impurity band. 

In summary, based on microscopic calculations, we constructed a
many-body Hamiltonian that is appropriate for describing ${\rm
Ga}_{1-x}{\rm Mn}_x {\rm As}$ in the very dilute limit, and estimated 
its parameters. We find that the
hopping of the carriers is strongly correlated with their spin.  This
spin-dependent hopping is crucial for capturing spin-orbit coupling
induced random anisotropy terms, or the lifetime of the magnon
excitations. Our calculations support the presence
of an impurity band for $x_{\rm active} \lesssim 0.015$ active Mn concentration.

\begin{acknowledgments}
We thank L. Brey, M. Berciu, S. Das Sarma, H.-H. Lin and C. Timm for
stimulating discussions. This work was supported by NSF PHY-9907949,
DMR-0233773, the NSF-MTA-OTKA Grant No. INT-0130446, Hungarian grants
No. OTKA T038162, T046303, T046267, the EU RTN HPRN-CT-2002-00302, the
Sloan foundation, and the Packard Foundation.
\end{acknowledgments}

\appendix

\section{Expressions for Angular Dependence of Induced Hole Polarization}
\label{app:hole_polarization}

With the wavefunctions (\ref{eq:Phi})  in hand, we can calculate
the average hole spin density around an isolated  Mn 
impurity, $\langle \vec j(r,\theta,\phi)\rangle$,  which reflects
the partial $d$-wave character of the Baldereschi-Lipari
wavefunctions. As an example, consider $\langle
j_z(r,\theta,\phi)\rangle_{F_z=3/2}$. Using the angular momentum 
addition rules we can express the orbital parts of the 
wave functions in Eq.~(\ref{eq:Phi}) as
\bea
&&|L=0,j=\frac32,F=\frac32,F_z=\frac32\rangle 
\to  Y_0^0 \; |\frac32 \rangle , 
\\
&&|L=2,j=\frac32,F=\frac32,F_z=\frac32\rangle \to 
%=
\\
&& \phantom{nn}
\to \sqrt{{2\over 5}} \;Y_2^2 \;|-\frac12\rangle
- \sqrt{{2\over 5}}\;Y_2^1 \;|\frac12\rangle +
\sqrt{{1\over 5}}\; Y_2^0 \; |\frac32\rangle\;,
\nonumber 
\eea
where the $Y_l^m$ denote the spherical functions, and 
the second terms  stand for the spin part of the $j=3/2$ 
wave function. 
Thus the full wave function reads
\begin{eqnarray}
&&\Phi_{F_z=3/2}=\left (f_0(r) Y_0^0(\theta,\phi) 
+{g_0(r)\over\sqrt{ 5}}Y_2^0(\theta,\phi)\right)
\Bigl|\frac32\Bigr\rangle \label{eq:decomposition_2}  \nonumber \\
&&\phantom{nn} 
- g_0(r)\sqrt{{2\over 5}}Y_2^1(\theta,\phi)\Bigl|\frac12\Bigr\rangle
+g_0(r)\sqrt{{2\over 5}}Y_2^2(\theta,\phi)\Bigl|-\frac12\Bigr\rangle\;.
\nonumber \\
\end{eqnarray}
Taking the expectation value of $j_\| \equiv j_z$ in this state gives, 
along with
analogous calculations in the other states and for the perpendicular 
component of the spin,  $j_\perp \equiv \cos(\phi)\; j_x + \sin(\phi)\;j_y$, 
\begin{widetext}
\bea
\langle  j_{\|}(\vec r)\rangle_{F_z=\pm 3/2} & =& \pm{3 \over 8 \pi}\left [f_0(r)^2+
f_0(r)g_0(r) (3 {\rm cos}^2(\theta)-1)+g_0(r)^2 {\rm cos}^4(\theta)\right]\;,
\\
\langle  j_{\bot}(\vec r)\rangle_{F_z=\pm 3/2} & = &
\pm{3 \over 4 \pi}\left [(f_0(r)+
{g_0(r)\over 2} (3 {\rm cos}^2(\theta)-1))g_0(r)+g_0(r)^2 {\rm sin}^2(\theta)\right]{\rm sin}(\theta){\rm cos}(\theta)\;,
\\
\langle  j_{\|}(\vec r)\rangle_{F_z=\pm 1/2}&=&\pm{1 \over 8 \pi}\Biggl [f_0(r)^2-
f_0(r)g_0(r) (3 {\rm cos}^2(\theta)-1)
+ {g_0(r)^2\over8}(5+12 {\rm cos}(2 \theta)-9 {\rm cos}(4\theta))\Biggr]\;,
\\
\langle  j_{\bot}(\vec r)\rangle_{F_z=\pm 1/2}&=&
\pm{3 \over 4 \pi}\Biggl [(f_0(r)-
{g_0(r)\over 2} (3 {\rm cos}^2(\theta)-1))g_0(r)\Biggr]{\rm sin}(\theta){\rm cos}(\theta) \;.
\eea
\end{widetext}

\section{Two-ion Problem}
\label{app:2ion}

Here we derive the parameters of the effective Hamiltonian
(\ref{eq:two_impurity}) using the molecular orbitals for a pair of Mn
ions.\cite{durst} Since the exchange interaction with the Mn core is
much less than the binding energy of the holes, and the on-site
interaction energy, we neglect its effect on the parameters of the
effective Hamiltonian (\ref{eq:H_2ion}).  The local field created by
the Mn core spin on the holes is later treated self-consistently in a
mean field and variational calculation described in
Sec.~\ref{sec:meanfield}.

\begin{figure}[h]
\begin{center}
\epsfig{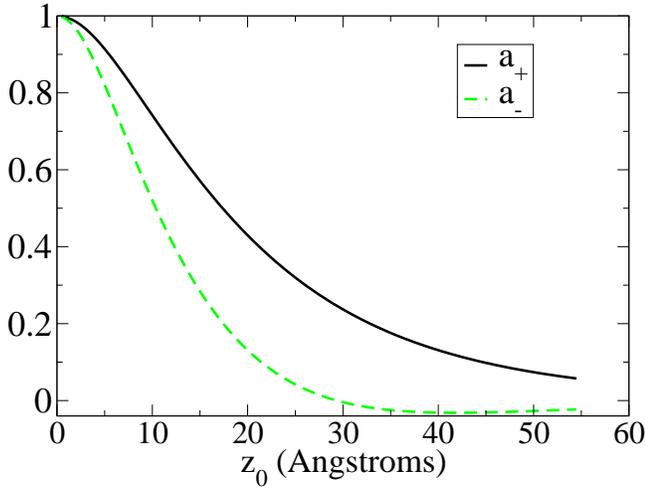}
\end{center}
\vskip0.1cm \caption{\label{fig:ab} (Color online.) Wave function overlaps for the two
states on site 1 and site 2. The overlaps are computed from Eq.~(B5).}
\end{figure}

\begin{figure}[h]
\begin{center}
\epsfig{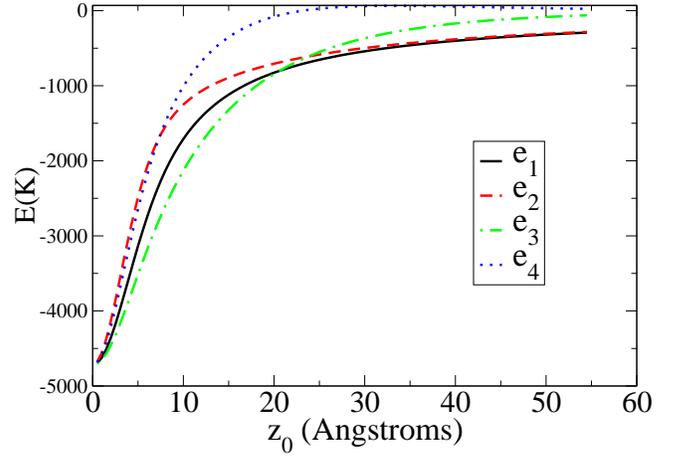}
\end{center}
\vskip0.1cm \caption{\label{fig:ei} (Color online.) Interaction overlaps for the interactions
defined in the text, Eqs.~(B6) and (B7).}
\end{figure}

We solve the problem in the eight-dimensional subspace spanned 
by the $F=3/2$ acceptor states centered on
each impurity obtained through the variational calculations of
Sec.~\ref{sec:variational_calculation}.  
As we discussed in the main text, within the spherical approximation
used throughout this paper, $F_z$ is conserved
if the two impurities are aligned along the $z$ axis.
In this case the sectors of different $F_z$ decouple. Furthermore, 
because of time reversal symmetry, the overlap matrices $S$ (see 
Sec.~\ref{sec:variational_calculation}) 
and Hamiltonian matrix elements  are identical for $F_z = \pm 3/2$ and 
for $F_z= \pm 1/2$. In the $F_z = \pm 3/2$ sector these are given by
\begin{equation}
S^{(3/2)}=\pmatrix{1 & a_+ \cr
  a_+ &1 \cr}
\end{equation}
and 
\begin{equation}
H^{(3/2)}=
\pmatrix{   E_0+e_1& e_3+a_+ \; E_0 \cr 
  e_3+ a_+\; E_0 & E_0+e_1\cr}\;,
\end{equation}
while for the $F_z=\pm 1/2$ subspace we have 
\bea
S^{(1/2)}&=&\pmatrix{1 & a_- \cr
  a_- &1 \cr}\;
\\
H^{(1/2)}&=&
\pmatrix{E_{0}+e_2& e_4+a_-\; E_{0} \cr
e_4+a_-\; E_{0} & E_{0}+e_2\cr}\;.
\eea
The two columns of these matrices correspond to the two Mn sites, 
and the constants $a_\pm$,  and $e_{1},\dots e_4$ denote various matrix 
elements
between the wavefunction of a hole at site 1 and a hole at site 2. 
The explicit formulas for these
quantities are given below. $E_{0}$ is the ground state energy of
the single bound hole as determined in
Sec.~\ref{sec:variational_calculation}.  Using Eq.~(\ref{eq:Phi}),
expanding the angular parts in spherical harmonics and then rewriting
the expressions in cylindrical coordinates, we have $r=\sqrt{\rho^2+z^2}$,
with $\rho$ the radial coordinate.  To simplify our expressions, we
introduce the notations $f_0\equiv f_0(r(\rho, z))$ (likewise for $g_0$),
$\tilde z \equiv z - z_0$,  $\tilde r \equiv \sqrt{\rho^2+(z-z_0)^2}$, and  
${\tilde f}_0 = f_0(\tilde r)$ (and likewise for ${\tilde g}_0 \equiv 
g_0(\tilde r)$), with $z_0$ the distance between the two impurities, 
and express the above matrix elements as
\begin{widetext}
\begin{eqnarray}
a_\pm &= & \int_0^\infty \rho d\rho \int_{-\infty}^\infty dz \Biggl[  {1\over 2}
\left ( f_0 \pm {g_0\over 2}\left (3 {z^2\over r^2}-1\right )\right ) 
\left ({\tilde f}_0 \pm
{{\tilde g}_0 \over 2}\left (3 {{\tilde z}_0^2\over {\tilde r}^2}-1\right )\right )
+{3\over 2}\;g_0\; {\tilde g}_0 \;
{\rho^2\over r^2 {\tilde r}^2} \left(z\tilde z +{1\over 4}
\rho^2\right)\Biggr]\;,
\\
e_{1,2}& =& \int_0^\infty \rho d\rho \int_{-\infty}^\infty dz \;V_2(\tilde r)\; \Biggl[  {1\over 2}
\left ( f_0 \pm  {g_0\over 2}\left (3 {z^2\over r^2}-1\right )\right )^2 +{3\over 2}g_0^2{\rho^2\over r^4 }\left(z^2+{1\over 4}
\rho^2\right)\Biggr] \;,
\\
e_{3,4} &=&  \int_0^\infty \rho d\rho \int_{-\infty}^\infty dz \; V_2(\tilde r) \; \Biggl[  {1\over 2}
\left ( f_0 \pm {g_0\over 2}\left (3 {z^2\over r^2}-1\right )\right )
\left ({\tilde f}_0 \pm
{{\tilde g}_0 \over 2}\left (3 {{\tilde z}_0^2\over {\tilde r}^2}-1\right )\right)
+{3\over 2}\;g_0\; {\tilde g}_0 \;
{\rho^2\over r^2 {\tilde r}^2} \left(z\tilde z +{1\over 4}
\rho^2\right)\Biggr], 
\end{eqnarray}
\end{widetext}
where $V_2 \equiv V_1(\tilde r)$ is given by 
Eq.~(\ref{eq:V1}). 
It should be kept in mind that $a_\pm$, the hole binding energy $E_{0}$ 
and the four $e_i$ all depend on the spherical spin-orbit strength $\mu$, and must be evaluated
numerically.  These parameters are shown in Figs.~\ref{fig:ab} and \ref{fig:ei}.  
Having these parameters at hand, we can simply determine the effective 
parameters $t_{1/2}$, $t_{3/2}$, $K$, and $E$ in Eq.~(\ref{eq:two_impurity})
by equating the spectrum of the two Mn ions with that of the effective Hamiltonian 
Eq.~(\ref{eq:two_impurity}). In this way we obtain
\bea
t_{3/2}&=&\frac{a_+ e_1-e_3}{a_+^2-1}\;,
\\
t_{1/2}& =&\frac{a_- e_2-e_4}{a_-^2-1}
\;,
\\
K &=& \frac12 \left({a_+ e_3 - e_1\over a_+^2 - 1}
- {a_- e_4 - e_2\over a_-^2 - 1}\right)\;,
\\
E &=& \frac12 \left({a_+ e_3 - e_1\over a_+^2 - 1}
+ {a_- e_4 - e_2\over a_-^2 - 1}\right)\;.
\eea
These parameters have been plotted in Fig.~\ref{fig:couplings}.


\section{Derivation and Evaluation of On-site Interactions $U_N$ and $U_F$}
\label{app:onsite_evaluation}

In the dilute limit it is important to include the effects of interactions
between holes.  Here we only consider the on-site interaction of the 
holes which dominate all other interactions
due to the localized nature of the molecular orbitals.

In second quantized form the interaction between two holes is
\begin{equation}
\hat H_{\rm on-site}=\frac{1}{2} 
\sum_{f_1,f_2,f_3,f_4} U_{f_1,f_2,f_3,f_4} c^\dagger_{f_1}
c^\dagger_{f_2}c_{f_3} c_{f_4}\;,
\label{eq:star} 
\end{equation}
where $U_{f_1,f_2,f_3,f_4}$ denotes the usual Coulomb integral

\begin{eqnarray}
U_{f_1,f_2,f_3,f_4}=\sum_{\mu,\nu} \int d^3\vec{r} \int d^3\vec{r}' 
\Phi^*_{f_1}(\vec{r},\mu) \Phi_{f_4}(\vec{r},\mu) \nonumber \\
\times {e^2 \over \epsilon \vert \vec{r} - \vec{r}' \vert } 
\Phi^*_{f_2}(\vec{r}^\prime,\nu) \Phi_{f_3}(\vec{r}^\prime,\nu)\;,
\label{eq:interaction_integrals}
\end{eqnarray}
and where we have again restricted ourselves to the same $F=3/2$ 
subspace, and correspondingly 
 the eigenvalues of the z-component of $F_z$, $f_i$, may take on 
the values $\pm 3/2$ and $\pm 1/2$. Here $\mu,\nu$ are the eigenvalues
of $j_z$.  
The wave functions $\Phi_{f_i}(\vec{r},\mu)$ have been determined previously 
 with the variational calculation outlined in
Sec~\ref{sec:variational_calculation}. 
(See Eq.~(\ref{eq:Phi}) and
Eq.~(\ref{eq:decomposition_2}) for an
illustration of how the angular dependence of
$\Phi_{f_i}(\vec{r},\mu)$ is obtained. A simple projection of
$\langle j=3/2,j_z=\mu=1/2 \vert$ into Eq.~(\ref{eq:decomposition_2}) picks
out $\Phi_{3/2}(\vec{r},1/2)=-g_0(r)\sqrt{{2\over 5}}Y_{2}^1(\theta,\phi)$, 
for example.)

Fortunately, we do not have to compute all these matrix elements
if we rewrite   Eq.~(\ref{eq:star}) in terms of {\em two-hole} scattering 
processes and  exploit rotational symmetry. 
Two holes can only take an  $F=0$ or an  $F=2$ configuration
within the ground state multiplet because of the Pauli principle. 
One can verify by direct evaluation that the $F=F_z = 2$ two-hole state 
is created by the following operator from the vacuum,
\begin{equation}
\hat D_2=c^\dagger_{1/2}\;c^\dagger_{3/2}\;,
\end{equation}
while the $F=2$ states of lower $F_z$ can 
be produced by applying the lowering operator. The corresponding operators
read: 
\begin{eqnarray}
\hat D_1 &=& c^\dagger_{-1/2}\;c^\dagger_{3/2}\;,\\
\hat D_0 &=& \frac{1}{\sqrt 2} (c^\dagger_{-3/2}\;c^\dagger_{3/2}+
c^\dagger_{-1/2}\;c^\dagger_{1/2})\;, \\
\hat D_{-1} &=& c^\dagger_{-3/2}\;c^\dagger_{1/2}\;,\\
\hat D_{-2} & = &c^\dagger_{-3/2}\;c^\dagger_{-1/2}\;.
\end{eqnarray}
Likewise for the sole $F=0$ operator we get,
\begin{equation}
\hat S_0 = \frac{1}{\sqrt 2} (c^\dagger_{-1/2}\;c^\dagger_{1/2}-
c^\dagger_{-3/2}\;c^\dagger_{3/2})\;.
\end{equation}

Since these operators transform as $F=0$ and $F=2$ tensor operators 
under SU(2) rotations, the interaction Hamiltonian must have the form
\begin{equation}
\hat H_{\rm on-site} =  U_D \sum_{m} \hat D^\dagger_m \hat D_m +  U_S \hat S^\dagger_0 \hat S_0\;.
\label{eq:irreducible_hamiltonian}
\end{equation}
We can, however, use instead of the decomposition above 
the following two SU(2) invariants too:
\be
\hat H_{\rm int} = \frac{U_N}{2}  :{\hat N}^2: + \,\frac{U_F}{2} 
 :\vec{F}^2:\;,
\label{eq:onsite_hamiltonian}
\ee
where $:\dots:$ denotes normal ordering and
$\hat N = \sum_{f} c^\dagger_{f} c_{f}$, and
$ \hat {\vec{F}} = \sum_{f,f'} c^\dagger_{f} \vec{F}_{f,f'} c_{f'}$
denote the number of holes and their total spin operator. 
It is easy to determine the relation of the constants 
$U_F$ and $U_N$ to $U_D$ and $U_S$
if one rewrites Eq.~(\ref{eq:onsite_hamiltonian}) using the identities
\bea
:{\hat N}^2: & = & {\hat N}^2 - {\hat N}\;,
\nonumber\\
 :\vec{F}^2: &  = &  \vec{F}^2 - \frac{15}4 {\hat N}\;,
\eea
and compares the action of Eq.~(\ref{eq:irreducible_hamiltonian})
and Eq.~(\ref{eq:onsite_hamiltonian}) on the $N=2$ and $F=0,2$ states.
This simple algebra gives:

%One can then write projection operators (with indicies suppressed) as
%\begin{equation}
%\hat P_D =  c^\dagger c^\dagger c c \frac{\vec{F}^2}{6}\;,
%\end{equation}
%and 
%\begin{equation}
%\hat P_S =  c^\dagger c^\dagger c c \left(1- \frac{\vec{F}^2}{6}\right )\;,
%\end{equation}
%where $\vec{F}=\vec{F_1}+\vec{F_2}$ is the total $\vec{F}$ of the two holes, so that
%\begin{equation}
%\frac{\vec{F}^2}{6} =\frac{5}{4} + \frac{\vec{F_1}\cdot \vec{F_2}}{3}\;.
%\end{equation}
%Hence, for any specified set of indicies 
%\begin{equation}
%U_D \hat P_D + U_S \hat P_S = \left (\frac{5 U_D - U_S}{4} + \frac{U_D-U_S}{3}\vec{F_1}\cdot \vec{F_2} \right )c^\dagger c^\dagger c c\;,
%\end{equation}
%where we can define 
\bea%\begin{equation}
U_N &=& \frac{5 U_D - U_S}{4}\;,
\label{eq:U_N}
\\
U_F &=& \frac{U_D-U_S}{3}\;.
\label{eq:U_F}
\eea

By comparing the matrix elements of
Eq.~(\ref{eq:irreducible_hamiltonian}) to the matrix elements of
Eq.~(\ref{eq:star}), we can evaluate  $U_D$ and $U_S$ in terms of
the $U_{f_1,f_2;f_4,f_3}\equiv U_{f_1,f_2,f_3,f_4}$ which in turn allow us to evaluate the $U_N$
and $U_F$ of Eq.~(\ref{eq:onsite_hamiltonian}).  Carrying out this
calculation, one obtains
\begin{eqnarray}
U_D&=&2\left( U_{\frac32,\frac12;\frac32,\frac12}-U_{\frac32,\frac12;\frac12,\frac32}\right)\\
U_S&=&4\left( U_{\frac12,-\frac12;\frac12,-\frac12}-U_{\frac12,-\frac12;-\frac12,\frac12}\right )-U_D\;,\nonumber \\
\end{eqnarray}
and therefore,
\bea 
U_N & = &3 \left(U_{\frac32,\frac12;\frac32,\frac12}- U_{\frac32,\frac12;\frac12,\frac32}\right )
\nonumber\\
& -&U_{\frac12,-\frac12;\frac12,-\frac12}+U_{\frac12,-\frac12;-\frac12,\frac12}
\\
%\end{eqnarray}
%\begin{eqnarray}
U_F&=&\frac{4}{3}\Bigl(U_{\frac32,\frac12;\frac32,\frac12}-U_{\frac32,\frac12;\frac12,\frac32}\nonumber \\
&-&U_{\frac12,-\frac12;\frac12,-\frac12}+U_{\frac12,-\frac12;-\frac12,\frac12}\Bigr )\;.
\end{eqnarray}

To obtain a numerical value of $U_N$ and $U_F$ we must determine the
matrix elements $U_{3/2,1/2;3/2,1/2}$, $U_{3/2,1/2;1/2,3/2}$,
$U_{1/2,-1/2;1/2,-1/2}$ and $U_{1/2,-1/2;-1/2,1/2}$ by evaluating the
integrals in Eq.~(\ref{eq:interaction_integrals}).  These integrals
depend on the radial wave functions that we evaluated variationally in
Sec.~\ref{sec:variational_calculation} and are material (parameter)
specific.  In order to evaluate the integrals in
Eq.~(\ref{eq:interaction_integrals}) the $\Phi_{\mu,f_i}(\vec{r})$
must be decomposed into spherical harmonics. Various products of
spherical harmonics appear in the integrand.  The integrals can be evaluated
by making use of the important formula
\begin{equation}
{1\over \vert \vec{r} -\vec{r}'\vert}=\frac{4 \pi}{r_>}\sum_{l=0}^\infty\sum_{m=-l}^l \left(\frac{r_<}{r_>}\right)^l \frac{(-1)^m}{2l+1}{Y_{l}^m}^*(\Omega)Y_{l}^m(\Omega')\;,
\end{equation}
where $\Omega$ ($\Omega'$) is the angle of $\vec{r}$ ($\vec{r}'$). Here
$r_>$ ($r_<$) is the greater (lesser) of $r$ and $r'$.  With this
formula, most of the integrals vanish and the few remaining integrals
yield
\bea
U_{\frac32,\frac12;\frac32,\frac12} &=& \frac{e^2}{\epsilon a_{\rm eff}}\; (I_1 - I_2)\;,
\\ \nonumber \\
U_{\frac32,\frac12;\frac12,\frac32} & = &\frac{e^2}{\epsilon a_{\rm eff}} \; 2 I_2\;,
\\
U_{\frac12,-\frac12;\frac12,-\frac12}&=& \frac{e^2}{\epsilon a_{\rm eff}} \;(I_1 + I_2)\;,
\\
U_{\frac12,-\frac12;-\frac12,\frac12}&=&0\;,
\eea
where the prefactor gives the energy scale of the interaction,
\begin{equation}
\frac{e^2}{\epsilon a_{\rm eff}}=31.6 \; {\rm meV}\;,
\end{equation}
and $I_1$ and $I_2$ denote the following integrals:
%\begin{widetext}
\bea
I_1 &=& \int_0^\infty\!\!\! r^2 dr \int_0^\infty \!\!\!r'^2 dr' \frac{1}{r_>}
(f^2_0(r) + g^2_0(r))
\nonumber \\
&&\phantom{nnn}\times(f^2_0(r') + g^2_0(r'))\;,
\\
I_2 &=& \int_0^\infty\!\!\! r^2 dr \int_0^\infty \!\!\!r'^2 dr' 
\frac{4}{25}\frac{r_<^2}{r_>^3}  f_0(r)g_0(r)f_0(r')g_0(r')
\;.\nonumber \\
\eea
%\begin{eqnarray}
%U_{3/2,1/2;3/2,1/2}=\frac{e^2}{\epsilon a_{\rm eff}}\int_0^\infty\!\!\! r^2 dr \int_0^\infty \!\!\!r'^2 dr' \frac{1}{r_>}\Biggl[ (f_0(r)^2 + g_0(r)^2)(f_0(r')^2 + g_0(r')^2)-\frac{4}{25}\left(\frac{r_<}{r_>}\right)^2  f_0(r)g_0(r)f_0(r')g_0(r')\Biggr]\;,\nonumber \\
%\end{eqnarray}
%\begin{eqnarray}
%U_{3/2,1/2;1/2,3/2}=\frac{e^2}{\epsilon a_{\rm eff}} \int_0^\infty r^2 dr \int_0^\infty r'^2 dr' \frac{8}{25}\frac{1}{r_>}\left(\frac{r_<}{r_>}\right)^2  f_0(r)g_0(r)f_0(r')g_0(r')\;,
%\end{eqnarray}
%\begin{eqnarray}
%U_{1/2,-1/2;1/2,-1/2}=\frac{e^2}{\epsilon a_{\rm eff}}\int_0^\infty \!\!\!\!\!r^2 dr \int_0^\infty \!\!\!\!\!r'^2 dr' \frac{1}{r_>}\Biggl[ (f_0(r)^2 + g_0(r)^2)(f_0(r')^2 + g_0(r')^2)+\frac{4}{25}\left(\frac{r_<}{r_>}\right)^2  f_0(r)g_0(r)f_0(r')g_0(r')\Biggr]\;.\nonumber \\
%\end{eqnarray}
%\end{widetext}
Evaluating these integrals one obtains $U_N=2570 \; {\rm K}$ and $U_F=-51 \;{\rm K}$.
%\begin{equation}
%U_N=2570 \; {\rm K}\;,
%\end{equation}
%and 
%\begin{equation}
%U_F=-51 \;{\rm K}\;.
%%%%%\end{equation}

%\bibliography{semiconductor}

\end{document}